\documentclass[aps,prd,twocolumn,nofootinbib]{revtex4}

\usepackage{graphicx}
\usepackage{color}
\usepackage[colorlinks=true, linkcolor=blue, citecolor=blue, urlcolor=blue]{hyperref}

\begin{document}

\title{Elimination of QCD Renormalization Scale and Scheme Ambiguities}

\author{Sheng-Quan Wang$^1$}
\email[email:]{sqwang@cqu.edu.cn}

\author{Stanley J. Brodsky$^2$}
\email[email:]{sjbth@slac.stanford.edu}

\author{Xing-Gang Wu$^3$}
\email[email:]{wuxg@cqu.edu.cn}

\author{Jian-Ming Shen$^4$}
\email[email:]{cqusjm@cqu.edu.cn}

\author{Leonardo Di Giustino$^{5,6}$}
\email[email:]{leonardo.digiustino@uninsubria.it}

\address{$^1$Department of Physics, Guizhou Minzu University, Guiyang 550025, P.R. China}
\address{$^2$SLAC National Accelerator Laboratory, Stanford University, Stanford, California 94039, USA}
\address{$^3$Department of Physics, Chongqing University, Chongqing 401331, P.R. China}
\address{$^4$School of Physics and Electronics, Hunan Provincial Key Laboratory of High-Energy Scale Physics and Applications, Hunan University, Changsha 410082, P.R. China}
\address{$^5$Department of Science and High Technology, University of Insubria, via valleggio 11, I-22100, Como, Italy}
\address{$^6$INFN, Sezione di Milano-Bicocca, 20126 Milano, Italy}

\begin{abstract}

The setting of the renormalization scale ($\mu_r$) in the perturbative QCD (pQCD) is one of the crucial problems for achieving precise fixed-order pQCD predictions. The conventional prescription is to take its value as the typical momentum transfer $Q$ in a given process, and theoretical uncertainties are then evaluated by varying it over an arbitrary range. The conventional scale-setting procedure introduces arbitrary scheme-and-scale ambiguities in fixed-order pQCD predictions. The principle of maximum conformality (PMC) provides a systematic way to eliminate the renormalization scheme-and-scale ambiguities. The PMC method has rigorous theoretical foundations; it satisfies the renormalization group invariance (RGI) and all of the self-consistency conditions derived from the renormalization group. The PMC has now been successfully applied to many physical processes. In this paper, we summarize recent PMC applications, including event shape observables and heavy quark pair production near the threshold region in $e^+e^-$ annihilation and top-quark decay at hadronic colliders. In addition, estimating the contributions related to the uncalculated higher-order terms is also summarized. These results show that the major theoretical uncertainties caused by different choices of $\mu_r$ are eliminated, and the improved pQCD predictions are thus obtained, demonstrating the generality and applicability of the PMC.

\end{abstract}

\maketitle

\section{Introduction}

The asymptotic freedom property of quantum chromodynamics (QCD) was proposed by Politzer, Gross, and Wilczek 50 years ago~\cite{Politzer:1973fx, Gross:1973id}. Due to the asymptotic freedom property, the strong interaction
whose magnitude can be characterized by the QCD strong coupling
$\alpha_s$ becomes small at very short distances, allowing
perturbative calculations for observables involving large momentum
transfer. The strong coupling $\alpha_s$ is scale-dependent, and
its behavior is controlled by renormalization group equation (RGE),
\begin{eqnarray}
\beta(\alpha_s)&=&\frac{d}{d\ln\mu^2_r}\left(\frac{\alpha_s(\mu_r)}{4\pi}\right)=-\sum^{\infty}_{i=0}\beta_i\left(\frac{\alpha_s(\mu_r)}{4\pi}\right)^{i+2}. \label{RGEquation}
\end{eqnarray}
The $\beta$ functions $\beta_0$, $\beta_1$, $\cdots$ are one-loop, two-loop, $\cdots$ corrections, respectively.

In the framework of perturbative QCD (pQCD), the prediction for an
observable $\rho$ at the $n_{\rm th}$-order level can be expressed
as a perturbative series over the QCD coupling $\alpha_s$, i.e.,

\begin{eqnarray}
\rho=\sum^{n}_{i=0}C_i\,\alpha_s(\mu_r)^{p+i}, \label{obsrho}
\end{eqnarray}
where $p$ is the power of the $\alpha_s(\mu_r)$ for the tree-level
terms. The scale $\mu_r$ represents the initial choice of
renormalization scale. The coefficients $C_1$, $C_2$, $\cdots$ are
one-loop, two-loop, $\cdots$ corrections, respectively.
The pQCD predictions, calculated up to all orders with
$n\to\infty$, are independent of the choice of the
renormalization scheme and renormalization scale because of the
renormalization group invariance (RGI). At any finite order, the
renormalization scheme and scale dependence of the QCD coupling
constant $\alpha_s(\mu_r)$ and of the QCD perturbative coefficients
$C_{i}$ only partially cancel. For example, it has been conventional
to guess the renormalization scale $\mu_r$, choosing
among the typical scales of a process, e.g., the typical momentum transfer $Q$,
in order to minimize large logarithmic corrections and
achieve relativistically more convergent series.
This conventional procedure breaks the RGI and introduces renormalization
scheme-and-scale ambiguities in pQCD predictions. The conventional
scale-setting method also has the negative consequence that the resulting
pQCD series suffers from a divergent renormalon $(\alpha_s^n
\beta_0^n n!)$ series~\cite{Beneke:1998ui} characteristic of a
non-conformal series at order $n$. Furthermore, theoretical uncertainties estimated
by simply varying the renormalization scale $\mu_r$ over an arbitrary range such as
$\mu_r\in[Q/2, 2Q]$ are clearly unreliable, since they are
only sensitive to the $\beta$-dependent non-conformal terms, not
the entire pQCD series. One cannot judge whether the slow convergence is an intrinsic property of pQCD series or is due to
the improper choice of renormalization scale $\mu_r$.

The conventional scale-setting procedure is also inconsistent with the well-known
{Gell-Mann}-Low (GM-L) method used in QED~\cite{GellMann:1954fq}. In
practice, the GM-L method shows that, by fixing the scale to the
correct momentum flow, one can reabsorb all the vacuum
polarization diagrams into the running coupling. Thus, the
renormalization scale-setting procedure in QED is void of any
ambiguity. A self-consistent scale-setting procedure should be
adaptable to both QCD and QED. In the limit of
$N_C\rightarrow0$~\cite{Brodsky:1997jk}, predictions for
non-Abelian QCD theory must agree analytically with predictions
for Abelian QED, and this also includes the renormalization
scale-setting procedure. Thus, the elimination of the ambiguities
in order to achieve precise pQCD predictions is crucial for
testing the standard model (SM) and for searching for new physics
beyond the SM.

The well-known Brodsky--Lepage--Mackenzie (BLM) method has been
suggested in Ref.~\cite{Brodsky:1982gc} and has been improved to
all orders as the principle of maximum conformality
(PMC)~\cite{Brodsky:2011ta, Brodsky:2012rj, Brodsky:2011ig,
Mojaza:2012mf, Brodsky:2013vpa} method. The PMC is the underlying
principle for the BLM method and provides a systematic all-orders way to
eliminate the renormalization scheme-and-scale ambiguities. This
method extends the BLM procedure unambiguously to all orders, to
all processes, and to all gauge theories. The PMC method meets all
the rigorous theoretical requirements, satisfying both the
RGI~\cite{Wu:2013ei, Wu:2014iba, Wu:2019mky} and the
self-consistency conditions derived from the renormalization
group~\cite{Brodsky:2012ms}. The PMC method reduces to the GM-L method in the Abelian limit.
A remarkable achievement of the PMC is that the resulting
scale-fixed predictions for physical observables are independent
of the choice of renormalization scheme---a key requirement of
RGI.

In 2017, the PMC single-scale method (PMCs)~\cite{Shen:2017pdu} was suggested, which is is equivalent to the multi-scale
method~\cite{Brodsky:2011ta, Brodsky:2012rj, Brodsky:2011ig,
Mojaza:2012mf, Brodsky:2013vpa} in the sense of perturbative
theory. The PMCs method effectively replaces the
individual PMC scales at each order derived by using the PMC
multi-scale method in the sense of a mean value theorem. In 2020, we
used an additional property of renormalizable SU(N)/U(1) gauge
theories~\cite{DiGiustino:2020fbk}, ``Intrinsic Conformality
(iCF)'', which underlies the scale invariance of physical
observables. It shows that the scale-invariant perturbative series
shows the intrinsic perturbative nature of a pQCD observable. In
2022, following the idea of iCF, we suggested a
novel single-scale-setting method under the PMC with the purpose
of removing the conventional renormalization scheme-and-scale
ambiguities~\cite{Yan:2022foz}. In Ref.~\cite{Yan:2022foz}, it has
been demonstrated that the two PMCs methods are equivalent to each other in accuracy.
This equivalence indicates that, by using the RGE for fixing the value of the
effective coupling, it is equivalent to requiring that each loop term
must satisfy the scale invariance simultaneously, and vice versa.
Thus, using the RGE provides a rigorous way to resolve conventional scale-setting ambiguities.

\section{A Mini-Review of the PMC Scale-Setting Method}

The scale evolution of $\alpha_s$ is described by the RGE as shown
by Equation~(\ref{RGEquation}), which can be used recursively to
establish the perturbative pattern of $\{\beta_i\}$-terms at each
order. The pQCD prediction for a physical observable $\rho$ can be
reorganized into the specific ``degeneracy''
pattern~\cite{Bi:2015wea} as follows:\vspace{-12pt}
\begin{widetext}
\begin{eqnarray}
\rho(Q) &=& r_{1,0}\,{\alpha(\mu_r)^p} + \left( r_{2,0} + p \beta_0 r_{2,1} \right)\,{\alpha(\mu_r)^{p+1}} + \left( r_{3,0} + p \beta_1 r_{2,1} + (p+1){\beta _0}r_{3,1} + \frac{p(p+1)}{2} \beta_0^2 r_{3,2} \right)\,{\alpha(\mu_r)^{p+2}} \nonumber\\
&& + \left( r_{4,0} + p{\beta_2}{r_{2,1}} + (p+1){\beta_1}{r_{3,1}} + \frac{p(3+2p)}{2}{\beta_1}{\beta_0}{r_{3,2}} + (p+2){\beta_0}{r_{4,1}} + \frac{(p+1)(p+2)}{2}\beta_0^2{r_{4,2}} \right.\\
&&\left. + \frac{p(p+1)(p+2)}{3!}\beta_0^3{r_{4,3}} \right)\,{\alpha(\mu_r)^{p+3}} + \cdots,\nonumber
\label{rhobefPMC}
\end{eqnarray}
\end{widetext}
where $\alpha=\alpha_s/4\pi$ and $Q$ represents the kinematic
scale at which the observable is measured. The coefficients
$r_{i,0(i=1,2,\cdots)}$ are conformal parts and
$r_{i,j(i>j\geq1)}$ are non-conformal ones. All the non-conformal
coefficients $r_{i,j(i>j\geq1)}$ are, in principle, functions of the scales $\mu_r$ and $Q$.

Following the PMC multi-scale procedures~\cite{Brodsky:2011ta,
Brodsky:2012rj, Brodsky:2011ig, Mojaza:2012mf, Brodsky:2013vpa},
all the non-conformal $\{\beta_i\}$-terms in
Equation~(\ref{rhobefPMC}) are systematically eliminated to fix the
correct magnitudes of QCD running couplings at each order (their
arguments are referred to as PMC scales); the resulting perturbative series then matches the corresponding conformal theory with
$\beta=0$, leading to scheme-independent predictions. The divergent
renormalon contributions are eliminated, and the convergence of
the perturbative series is in general greatly improved. This is the same principle used in QED where all $\{\beta_i\}$-terms derived from the vacuum polarization corrections of the photon
propagator are absorbed into the QED coupling. As in QED, the resulting PMC scales are physical in the sense that
they reflect the virtuality of the gluon propagators at a given
order, and that they set the active flavors $n_f$. More explicitly, after applying the PMC multi-scale method, the pQCD series for the physical observable $\rho$~becomes
\begin{eqnarray}
\rho(Q) &=& r_{1,0}\,{\alpha(Q_1)^p} + r_{2,0}\,{\alpha(Q_2)^{p+1}} + r_{3,0}\,{\alpha(Q_3)^{p+2}} \nonumber\\
&& + r_{4,0}\,{\alpha(Q_4)^{p+3}} + \cdots,
\label{rhoafterPMC}
\end{eqnarray}
where $Q_{i=1,2,3,4}$ are the PMC scales. Due to uncalculated
higher-order contributions, there are two kinds of residual scale
dependences~\cite{Zheng:2013uja}. The first kind of residual scale dependence is from the PMC scale itself because the PMC scale is a perturbative expansion series in $\alpha_s$. The second kind of
residual scale dependence is from the last terms of the pQCD approximant
because its magnitude cannot be determined. These
residual scale dependencies are distinct from the conventional
renormalization scale ambiguities and are suppressed due to the perturbative
nature of the PMC scale.

In order to suppress the residual scale dependence, which also
makes the PMC scale-setting procedures simpler and more easily
automatized, the PMCs method has been suggested in
Ref.~\cite{Shen:2017pdu}. The PMCs method provides a self-consistent
way to achieve precise $\alpha_s$ running behavior in both the perturbative and nonperturbative
domains~\cite{Deur:2017cvd, Yu:2021yvw}. After applying the PMCs method,
the pQCD prediction for the physical observable $\rho$ can be written as
\begin{eqnarray}
\rho(Q) &=& r_{1,0}\,{\alpha(Q_\star)^p} + r_{2,0}\,{\alpha(Q_\star)^{p+1}} + r_{3,0}\,{\alpha(Q_\star)^{p+2}} \nonumber\\
&& + r_{4,0}\,{\alpha(Q_\star)^{p+3}} + \cdots.
\label{rhoafterPMCs}
\end{eqnarray}

The single PMC scale $Q_\star$ is determined by requiring all
the non-conformal $\{\beta_i\}$-terms to vanish simultaneously and
can be regarded as the overall effective momentum flow of the
process. The PMCs method exactly removes the second kind of residual scale dependence,
and the first kind of residual scale dependence is highly suppressed.
The PMCs method eliminates the renormalization scheme-and-scale ambiguities
and satisfies the standard RGI~\cite{Wu:2019mky}.

Up to now, the PMC approach has been successfully applied to many physical
processes (see, e.g.,~\cite{Wu:2014iba, Wu:2015rga, Wu:2019mky} for reviews),
including the Higgs boson production at the
LHC, the Higgs boson decays to
$\gamma\gamma$, $gg$, and
$b\bar{b}$ processes, the top-quark pair production
at the LHC and Tevatron and its decay process~\cite{Meng:2022htg}, the semihard
processes based on the BFKL approach~\cite{Brodsky:1999je,
Hentschinski:2012kr, Zheng:2013uja, Caporale:2015uva}, the
electron--positron annihilation to hadrons~\cite{Mojaza:2012mf,
Brodsky:2013vpa, Wu:2014iba}, the hadronic $Z^0$ boson
decays, the event shapes in electron--positron annihilation~\cite{DiGiustino:2020fbk, Wang:2021tak}, the
electroweak parameter $\rho$~\cite{Wang:2014wua, Yu:2021ujk}, the
$\Upsilon(1S)$ leptonic decay~\cite{Shen:2015cta, Huang:2019frb}, and
the charmonium \mbox{production~\cite{Wang:2013vn, Sun:2018rgx,
Yu:2020tri}}. In addition, the PMC
provides a possible solution to the $B\to\pi\pi$
puzzle~\cite{Qiao:2014lwa} and the
$\gamma\gamma^*\rightarrow\eta_c$ puzzle~\cite{Wang:2018lry}. In
the following, we present some recent PMC applications and a way
of estimating unknown contributions from uncalculated higher-order
terms by using the PMC pQCD series.

\section{Applications}

\subsection{New Analyses of Event Shape Observables in $e^+e^-$ Annihilation }

Event shapes represent an ideal platform for high-precision tests of QCD (see e.g.,~\cite{PDG:2020}
for a summary from Particle Data Group). The experiments at LEP and at SLAC have
measured event shape distributions with very high precision,
especially those at the $Z^0$ peak~\cite{Heister:2003aj,
Abdallah:2003xz, Abbiendi:2004qz, Achard:2004sv, Abe:1994mf}. On
the theoretical side, the pQCD corrections to event shape
observables have been calculated up to the next-to-next-to-leading
order (NNLO)~\cite{Gehrmann-DeRidder:2007nzq,
GehrmannDeRidder:2007hr, Ridder:2014wza, Weinzierl:2008iv,
Weinzierl:2009ms, DelDuca:2016csb, DelDuca:2016ily}. Currently, one finds that
the main obstacle for achieving highly precise measurements of
$\alpha_s$ from event shape variables is given by theoretical
uncertainties, especially those related to the renormalization
scale ambiguities.

Comprehensive PMC analysis for event shapes in $e^+e^-$ annihilation and a novel method for the precise determination of
the QCD running coupling $\alpha_s(Q^2)$ are shown in Refs.~\cite{Wang:2019isi, Wang:2021tak}. Interested readers may turn to these
studies for more details. In this paper, we only present the
main PMC results for two fundamental event shapes: the thrust
($T$)~\cite{Brandt:1964sa, Farhi:1977sg} and the $C$-parameter
($C$)~\cite{Parisi:1978eg, Donoghue:1979vi}.

\begin{figure}[htb]
\centering
\includegraphics[width=0.40\textwidth]{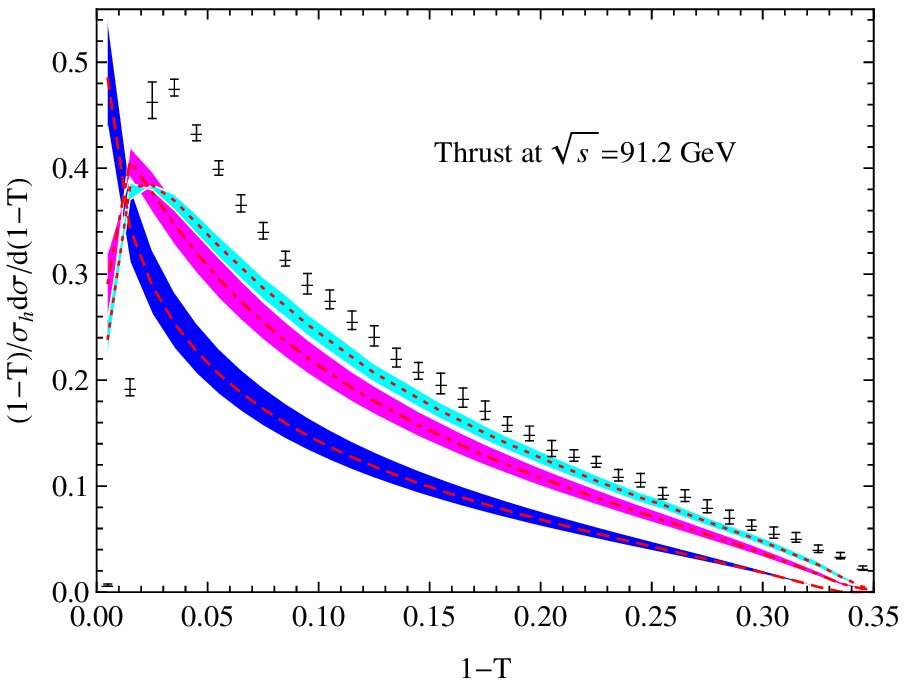}
\includegraphics[width=0.40\textwidth]{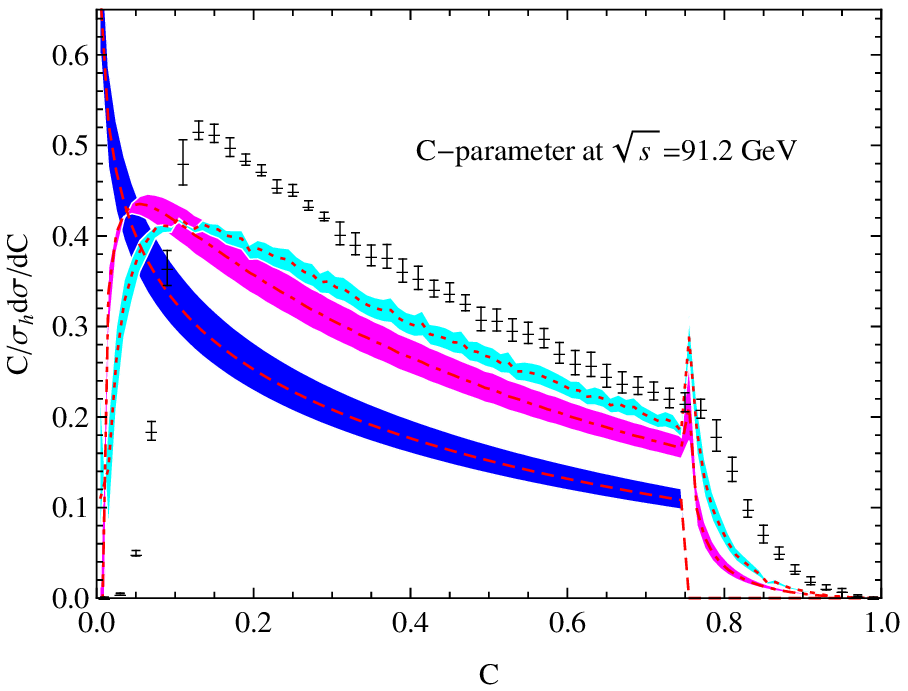}
\caption{The thrust ($T$) and $C$-parameter ($C$) distributions using the conventional scale-setting method at
$\sqrt{s}=91.2$ GeV, where the dashed, dot-dashed, and dotted lines
are the conventional results at LO, NLO, and
NNLO~\cite{GehrmannDeRidder:2007hr, Weinzierl:2009ms},
respectively. The bands are obtained by varying the scale $\mu_r\in[\sqrt{s}/2,2\sqrt{s}]$. The
experimental data are taken from the ALEPH
Collaboration~\cite{Heister:2003aj}.} \label{TCConVSdata}
\end{figure}

In the case of conventional scale setting, one simply sets the
renormalization scale $\mu_r$ to the center-of-mass energy
$\mu_r=\sqrt{s}$. We present the thrust and $C$-parameter
differential distributions using the conventional scale-setting
method at $\sqrt{s}=91.2$ GeV in Figure~\ref{TCConVSdata}. Results
show that even up to NNLO QCD corrections, the conventional results are plagued by large-scale uncertainty and
substantially deviate from the precise experimental data.

Moreover, the method does not improve the precision at higher orders,
since the results are totally arbitrary. In fact, varying the
$\mu_r\in[\sqrt{s}/2,2\sqrt{s}]$, the NLO calculation does not
overlap with the LO prediction, and the NNLO calculation does not
overlap with the NLO prediction. This indicates that the evaluation of
uncalculated higher-order (UHO) terms for event shape observables by
varying $\mu_r\in[\sqrt{s}/2,2\sqrt{s}]$ is not quantitatively
reliable. Worse, since the renormalization scale is simply set to $\mu_r=\sqrt{s}$,
only one value of $\alpha_s$ at the scale $\sqrt{s}$ can be extracted,
with an arbitrary large error given by the choice of the
renormalization scale $\mu_r$.

\begin{figure}[htb]
\centering
\includegraphics[width=0.40\textwidth]{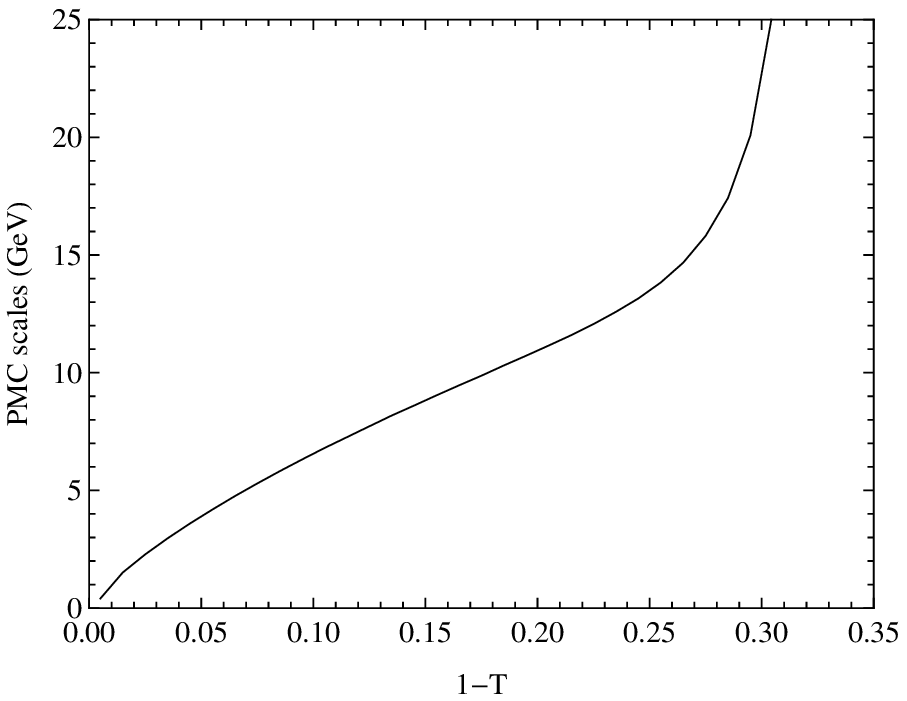}
\includegraphics[width=0.40\textwidth]{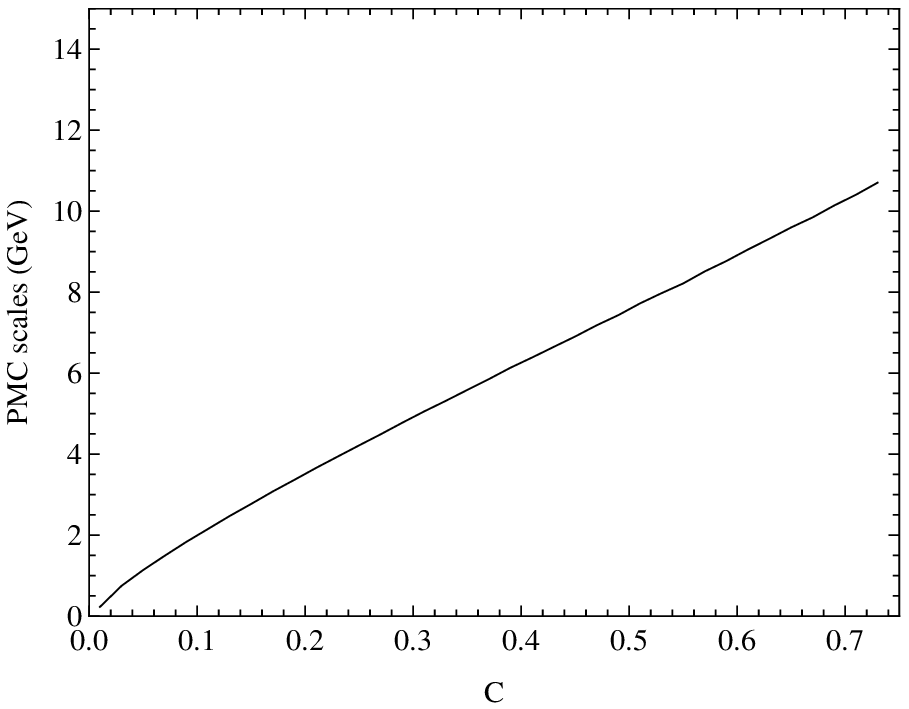}
\caption{{The PMC scales for the event shape observables thrust ($T$) and $C$-parameter ($C$) at \mbox{$\sqrt{s}=91.2$~GeV.} }}
\label{PMCscaleTC}
\end{figure}

The PMC scales are determined by absorbing all the $\beta$ terms of
the pQCD series. In Figure~\ref{PMCscaleTC}, we show the PMC scales for thrust and
$C$-parameter at the scale $\sqrt{s}=91.2$ GeV. The resulting PMC
scales are not a single value, but they monotonically increase with
the value of $T$ and $C$, reflecting the increasing virtuality of
the QCD dynamics. The number of active flavors $n_f$ changes with
the value of $T$ and $C$ according to the PMC scales. It is noted
that the quarks and gluons have soft virtuality near the two-jet
region (left boundary). As the argument of the $\alpha_s$
approaches the two-jet scale-region, the PMC scales are very soft.
Thus, the dynamics of the PMC scale reflect the correct physical
behavior when approaching the two-jet region. In addition, the PMC
scales are small in the wide kinematic regions compared to the
conventional choice of $\mu_r=\sqrt{s}$.

\begin{figure}[htb]
\centering
\includegraphics[width=0.40\textwidth]{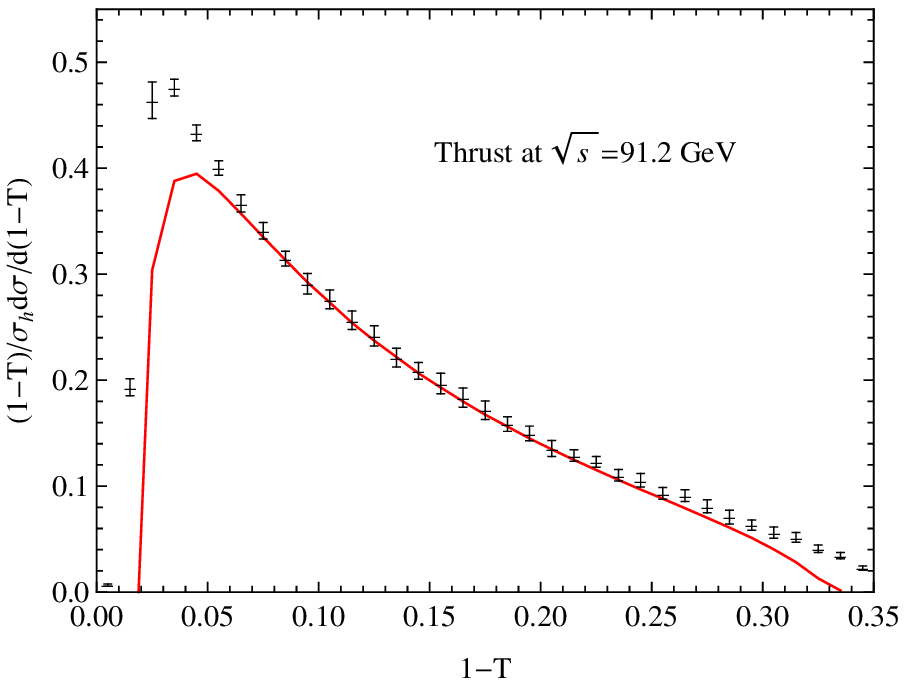}
\includegraphics[width=0.40\textwidth]{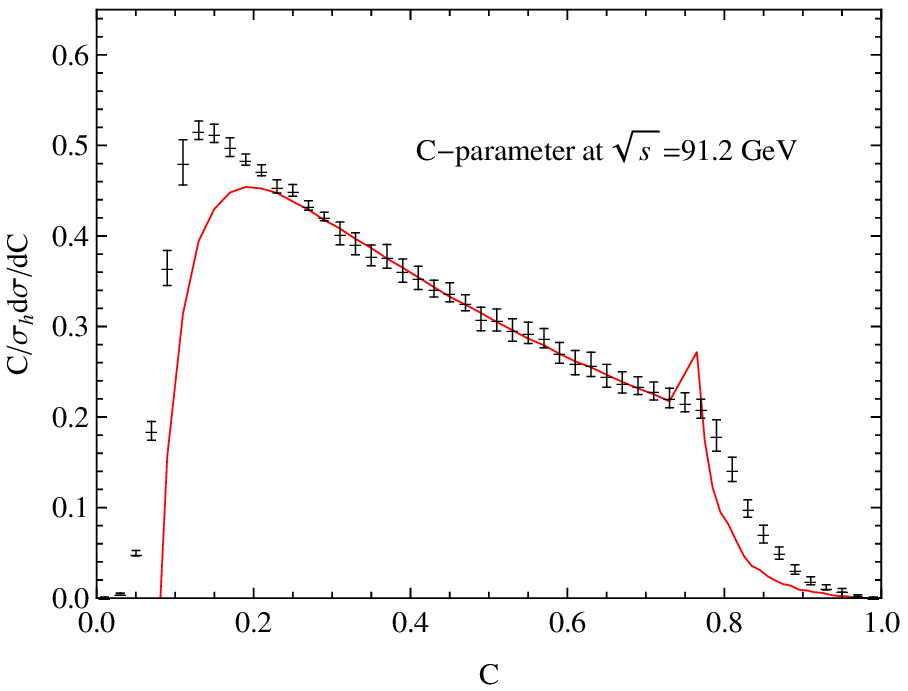}
\caption{The thrust ($T$) and $C$-parameter ($C$) distributions
using PMC scale setting for \mbox{$\sqrt{s}=91.2$ GeV.} The experimental
data are taken from the ALEPH
Collaboration~\cite{Heister:2003aj}.} \label{TCPMCVSdata}
\end{figure}

It is noted that the behavior of the PMC conformal coefficients is
significantly different from the pQCD terms given by the
conventional scale-setting method. Since the conformal
coefficients are renormalization scale-independent, the resulting
PMC predictions eliminate the renormalization scale uncertainty.
By setting all input parameters to their central values, we
present the thrust and $C$-parameter distributions using the PMC
scale-setting method for $\sqrt{s}=91.2$ GeV in
Figure~\ref{TCPMCVSdata}. This figure shows that the PMC predictions
improve for a wide range of values in the kinematic regions with
respect to the conventional scale-setting predictions and are in
excellent agreement with the experimental data, especially in the
intermediate kinematic regions. Since there are large logarithms
that spoil the perturbative regime of the QCD near the two-jet
region and there are missing higher-order contributions that are
important for the multi-jet region, the PMC predictions in these
regions show some deviations from experiments.

The resummation of large logarithms is thus required for the PMC
predictions, especially near the two-jet region. In fact, the
resummation of large logarithms has been extensively studied in
the literature.

\begin{figure}[htb]
\centering
\includegraphics[width=0.40\textwidth]{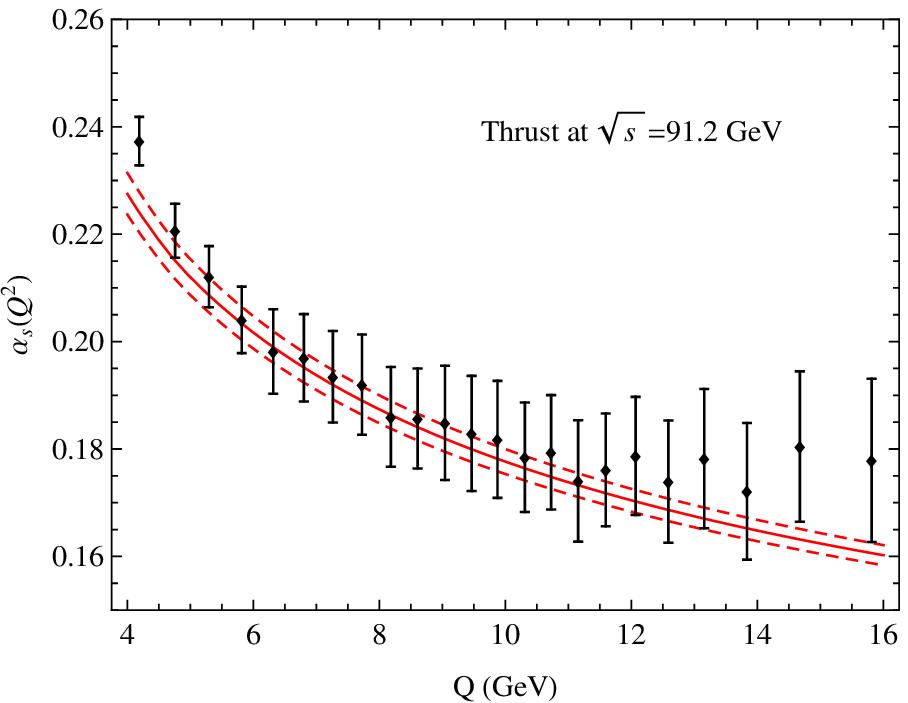}
\includegraphics[width=0.40\textwidth]{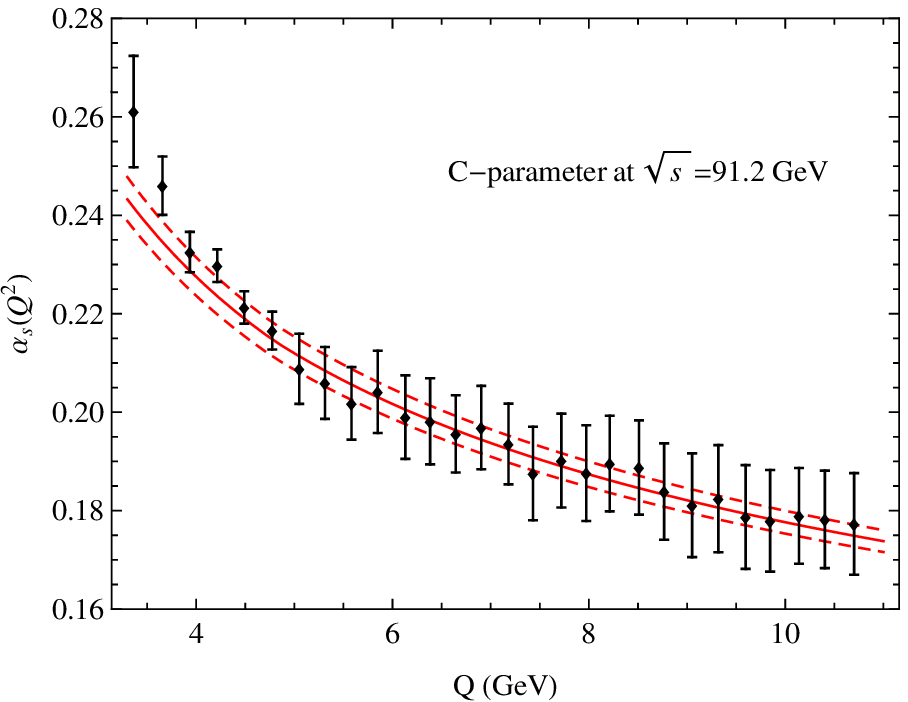}
\caption{The extracted running coupling $\alpha_s(Q^2)$ from the
thrust ($T$) and $C$-parameter ($C$) distributions by comparing
the PMC predictions with the ALEPH data~\cite{Heister:2003aj}
measured at a single energy of $\sqrt{s}=91.2$ GeV. As a
comparison, the three lines represent the world
average~\cite{PDG:2020}. } \label{alphasPMCTC}
\end{figure}

For the extraction of $\alpha_s$, since the renormalization scale
is simply set as $\mu_r=\sqrt{s}$ when using conventional scale
setting, only one value of $\alpha_s$ at scale $\sqrt{s}$ can be
extracted, as mentioned above. On the contrary, in applying the PMC method, since the PMC
scales vary with the value of the event shapes $T$ and $C$, we can
extract $\alpha_s(Q^2)$ over a wide range of $Q^2$ using the
experimental data at a single energy of $\sqrt{s}$. By comparing
PMC predictions with measurements at $\sqrt{s}=91.2$ GeV, we
present the extracted running coupling $\alpha_s(Q^2)$ from the
thrust and $C$-parameter distributions in Figure~\ref{alphasPMCTC}.


Figure~\ref{alphasPMCTC} shows that the extracted $\alpha_s(Q^2)$
in the ranges $4<Q<16$ GeV from the thrust and $3<Q<11$ GeV from
the $C$-parameter are in excellent agreement with the world
average evaluated using the value
$\alpha_s(M^2_Z)=0.1179$~\cite{PDG:2020} for the coupling at
$Z^0$ mass. Since the PMC method eliminates the renormalization
scale uncertainty, the extracted $\alpha_s(Q^2)$ is not plagued by
any uncertainty from the choice of the scale $\mu_r$. Thus, PMC
scale setting provides a remarkable way to verify the running of
$\alpha_s(Q^2)$ from event shape observables in $e^+e^-$ annihilation measured at a single energy $\sqrt{s}$.

The mean value of event shape observables
provides an important complement to the differential distributions
and to the determination of $\alpha_s$. The mean value of an event
shape $y$ is defined as
\begin{eqnarray}
\langle y\rangle &=&\int_0^{y_0}\frac{y}{\sigma_{h}}\frac{d\sigma}{dy}dy,
\end{eqnarray}
where $y_0$ is the kinematically-allowed upper bound of the $y$
variable, and the integration is over the full phase space.

In the case of a conventional scale setting, the mean values of $T$ and $C$
are plagued by the renormalization
scale uncertainties and substantially deviate from the
measurements even at NNLO~\cite{GehrmannDeRidder:2009dp,
Weinzierl:2009yz}, similar to the case of the differential
distributions.

Currently, the most common way to calculate the integral for the
mean values is to distinguish the two perturbative and
non-perturbative contributions and to calculate them separately.
This is known and extensively studied in the literature.
Nevertheless, some artificial parameters and a theoretical model have
to be introduced in order to match theoretical predictions with
experimental data.

After applying the PMC, we obtain
\begin{eqnarray}
\mu^{\rm{pmc}}_r|_{\langle1-T\rangle} = 0.0695\sqrt{s}
\end{eqnarray}
for the mean value of the thrust and
\begin{eqnarray}
\mu^{\rm{pmc}}_r|_{\langle C\rangle} = 0.0656 \sqrt{s}
\end{eqnarray}
for the mean value of the $C$-parameter. The PMC scales satisfy
$\mu^{\rm pmc}_r\ll \sqrt{s}$, reflecting the soft virtuality of
the underlying QCD subprocesses. We note that in the analysis of
Ref.~\cite{Heister:2003aj}, using a conventional scale setting leads
to an anomalously large value of $\alpha_s$, demonstrating again
that the correct description for the mean values requires
$\mu_r\ll \sqrt{s}$. The PMC scales for the differential
distributions of the thrust and $C$-parameter are also very small.
PMC scale setting is self-consistent with the differential
distributions of the event shapes and their mean values.

\begin{figure}[htb]
\centering
\includegraphics[width=0.40\textwidth]{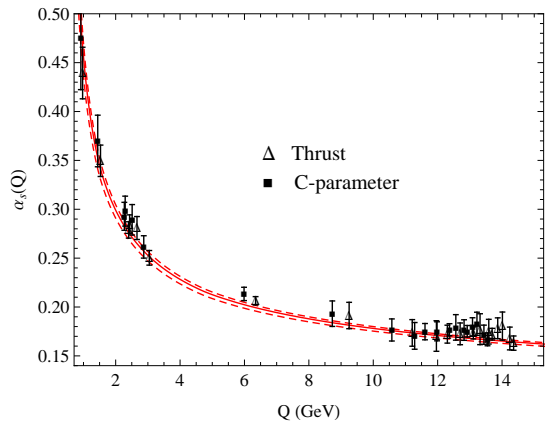}
\caption{The extracted QCD coupling $\alpha_s(Q^2)$ from the
thrust and $C$-parameter mean values by comparing PMC predictions
with the JADE and OPAL data~\cite{Abbiendi:2004qz, Pahl:2007zz}.
The error bars are the squared averages of the experimental and
theoretical errors. The three lines are the world
average~\cite{PDG:2020}. } \label{momentas}
\end{figure}

After using PMC scale setting, the thrust and $C$-parameter mean
values are increased, especially at small $\sqrt{s}$. The
scale-independent PMC predictions are in excellent agreement with
the experimental data over a wide range of center-of-mass energies
$\sqrt{s}$~\cite{Wang:2019isi}. Since we can obtain a high degree
of consistency between the PMC predictions and the measurements,
the QCD coupling $\alpha_s(Q^2)$ can be extracted with high
precision. The extracted QCD coupling $\alpha_s(Q^2)$ in the
$\overline{\rm MS}$ scheme from the thrust and $C$-parameter mean
values is presented in Figure~\ref{momentas}. This figure shows
that the extracted $\alpha_s(Q^2)$ values are mutually compatible and are
in excellent agreement with the world average.

A highly precise determination of the value of $\alpha_s(M^2_Z)$
fitting the PMC predictions to the measurements is achieved.
Finally, we obtain~\cite{Wang:2019isi}
\begin{eqnarray}
\alpha_s(M^2_Z) = 0.1185\pm0.0012
\end{eqnarray}
from the thrust mean value and
\begin{eqnarray}
\alpha_s(M^2_Z) = 0.1193^{+0.0021}_{-0.0019},
\end{eqnarray}
from the $C$-parameter mean value. Since the dominant renormalization scale
$\mu_r$ uncertainty is eliminated and the convergence of pQCD
series is greatly improved after using the PMC method, the precision of
the extracted $\alpha_s$ values is largely improved.

\subsection{Heavy Quark Pair Production in $e^+e^-$ Annihilation near the Threshold Region}

Heavy fermion pair production in $e^+e^-$ annihilation is a
fundamental process for the SM and of considerable interest for
other phenomena. Heavy quark interaction in the threshold region
is of particular interest due to the presence of singular terms
from the QCD Coulomb corrections. Physically, the renormalization
scale that reflects the subprocess virtuality becomes very soft
in this region. It is conventional to set the renormalization
scale to the mass of the heavy fermion $\mu_r=m_f$. This
conventional procedure obviously violates the physical behavior of
the QCD corrections and leads to results affected by systematic
errors, due to inherent scheme and scale uncertainties, and
predictions are quite unreliable in this kinematic region.

The quark pair production cross-section for $e^{+}e^{-}\rightarrow\gamma^{*}\rightarrow Q\bar{Q}$ at the two-loop level can be written as
\begin{eqnarray}
\sigma=\sigma^{(0)}\left[1 + \delta^{(1)}\,a_s(\mu_r) + \delta^{(2)}(\mu_r)\,a^2_s(\mu_r) + {\cal O}(a^3_s)\right],
\label{sigma:1}
\end{eqnarray}
where $a_s(\mu_r)={\alpha_s(\mu_r)}/{\pi}$. The LO cross section is
\begin{eqnarray}
\sigma^{(0)}=\frac{4}{3}\frac{\pi\,\alpha_e^2}{s}N_c\,e^2_Q\frac{v\,(3-v^2)}{2},
\label{qqpairLOpmc}
\end{eqnarray}
where $\alpha_e$ is the fine structure constant, $N_c$ is the
number of colors, and $e_Q$ is the $Q$ quark electric charge. The
quark velocity $v$ is $v=\sqrt{1-{4\,m_Q^2}/{s}}$, where $s$ is
the center-of-mass energy squared and $m_Q$ is the mass of the
quark $Q$.

The one-loop correction coefficient $\delta^{(1)}$ is $\delta^{(1)}=C_F({\pi^2}/{2\,v}-4)$.
The two-loop correction coefficient $\delta^{(2)}$ can be conveniently split into terms proportional to different $SU(3)$ color factors,
\begin{eqnarray}
\delta^{(2)} &=& C_F^2\,\delta^{(2)}_A + C_F\,C_A\,\delta^{(2)}_{NA} \nonumber\\
&& + C_F\,T_R\,n_f\,\delta^{(2)}_L + C_F\,T_R\,\delta^{(2)}_H.
\end{eqnarray}

The terms $\delta^{(2)}_{A}$, $\delta^{(2)}_{L}$, and
$\delta^{(2)}_{H}$ are the same in Abelian and non-Abelian
theories; the term $\delta^{(2)}_{NA}$ only arises in the
non-Abelian theory. This process offers the opportunity to rigorously
rigorously the scale-setting method in the non-Abelian
and Abelian theories.

The cross-section given in Equation~(\ref{sigma:1}) can be written
indicating explicitly the $n_f$-dependent and $n_f$-independent
parts, i.e.,\vspace{-9pt}
\begin{widetext}
\begin{eqnarray}
\sigma &=&\sigma^{(0)}\left[1+\delta^{(1)}_h\,a_s(\mu_r) + \left(\delta^{(2)}_{h,in}(\mu_r)+\delta^{(2)}_{h,n_f}(\mu_r)\,n_f\right)\,a^2_s(\mu_r) \right. \nonumber\\
&&\left. + \left(\frac{\pi}{v}\right)\,\delta^{(1)}_{v}\,a_s(\mu_r) + \left(\frac{\pi}{v}\right)\,\left(\delta^{(2)}_{v,in}(\mu_r)+\delta^{(2)}_{v,n_f}(\mu_r)\,n_f\right)\,a^2_s(\mu_r)+ \left(\frac{\pi}{v}\right)^2\,\delta^{(2)}_{v^2}\,a^2_s(\mu_r) + {\cal O}(a^3_s)\right].
\label{eqpairNNLO}
\end{eqnarray}
\end{widetext}

Coefficients $\delta^{(1)}_v$, $\delta^{(2)}_v$, and
$\delta^{(2)}_{v^2}$ are for the Coulomb corrections, while
coefficients $\delta^{(1)}_h$ and $\delta^{(2)}_h$ are for the
non-Coulomb corrections. These coefficients have been determined
in Refs.~\cite{Czarnecki:1997vz, Beneke:1997jm,
Bernreuther:2006vp} for the $\overline{\rm MS}$ scheme. Due to
their proportional form to powers of $(\pi/v)$, Coulomb
corrections are enhanced in the threshold region. This implies
that the renormalization scale can be relatively soft in this
region. Therefore, the PMC scales must be determined separately for the
non-Coulomb and Coulomb corrections~\cite{Brodsky:1995ds,
Brodsky:2012rj}. When the quark velocity $v\rightarrow0$, the
Coulomb correction dominates the contribution for the production
cross-section.

Absorbing the non-conformal term $\beta_0=11/3\,C_A-4/3\,T_R\,n_f$
into the running coupling constant, as implemented in the PMC
procedure, we obtain
\begin{eqnarray}
\sigma &=& \sigma^{(0)}\left[1+\delta^{(1)}_h\,a_s(Q_h) + \delta^{(2)}_{h,\rm sc}(\mu_r)\,a^2_s(Q_h) \right. \nonumber\\
&&\left. + \left(\frac{\pi}{v}\right)\,\delta^{(1)}_v\,a_s(Q_v) + \left(\frac{\pi}{v}\right)\,\delta^{(2)}_{v,\rm sc}(\mu_r)\,a^2_s(Q_v)\right. \\
&&\left. + \left(\frac{\pi}{v}\right)^2\,\delta^{(2)}_{v^2}\,a^2_s(Q_v) + {\cal O}(a^3_s)\right].\nonumber
\label{eqpairNNLOpmc}
\end{eqnarray}

The PMC scales $Q_i$ are

\begin{eqnarray}
Q_i = \mu_r\exp\left[\frac{3\,\delta^{(2)}_{i,n_f}(\mu_r)}{2\,T_R\,\delta^{(1)}_i}\right],
\end{eqnarray}
and the coefficients $\delta^{(2)}_{i,\rm sc}(\mu_r)$ are
\begin{eqnarray}
\delta^{(2)}_{i,\rm sc}(\mu_r) = \frac{11\,C_A\,\delta^{(2)}_{i,n_f}(\mu_r)}{4\,T_R}+\delta^{(2)}_{i,in}(\mu_r),
\end{eqnarray}
where $i=h$ and $v$ stand for the non-Coulomb and Coulomb
corrections, respectively. At the present two-loop level, the
conformal coefficients and the PMC scales are independent of the
renormalization scale $\mu_r$. Thus, the resulting cross-section
in Equation~(\ref{eqpairNNLOpmc}) is void of renormalization scale
uncertainties.

The V-scheme defined by the interaction potential between two
heavy quarks~\cite{Appelquist:1977tw, Fischler:1977yf,
Peter:1996ig, Schroder:1998vy, Smirnov:2008pn, Smirnov:2009fh,
Anzai:2009tm, Kataev:2015yha, Kataev:2023sru}, $V(Q^2) =
-{4\,\pi^2\,C_F\,a^V_s(Q)}/{Q^2}$, provides a physically-based
alternative to the usual $\overline{\rm MS}$ scheme. As in the
case of QED, when the scale of the coupling $a^V_s$ is identified
with the exchanged momentum, all vacuum polarization corrections
are resummed into $a^V_s$. By using the relation between $a_s$ and
$a^V_s$ at the one-loop level, i.e.,
\begin{eqnarray}
a^V_s(Q) = a_s(Q) + \left({31\over36}C_A-{5\over9}T_R\,n_f\right)a^2_s(Q),
\label{Vscheme:as}
\end{eqnarray}
we can perform a change of scheme, from the $\overline{\rm MS}$
scheme to the V-scheme, for the quark pair production cross-section. The corresponding perturbative coefficients in
Equation~(\ref{eqpairNNLO}) for the V-scheme are given in
Ref.~\cite{Wang:2020ckr}. The predictions using the PMC eliminate
the dependence from the renormalization scheme; this is explicitly
displayed in the form of ``commensurate scale relations''
(CSR)~\cite{Brodsky:1994eh, Lu:1992nt}.

The PMC scales in the $\overline{\rm MS}$ scheme are
$Q_h=e^{(-11/24)}\,m_Q$ for the non-Coulomb correction and $Q_v
=2\,e^{(-5/6)}\,v\,m_Q$ for the Coulomb correction. In the
V-scheme, the scales are $Q_h =e^{(3/8)}\,m_Q$ for the non-Coulomb
correction and $Q_v =2\,v\,m_Q$ for the Coulomb correction. The
scale $Q_h$ stems from the hard virtual gluon corrections, and
the scale $Q_v$ originates from the final state Coulomb
re-scattering. As expected, the scale $Q_h$ is of the order $m_Q$,
whereas the scale $Q_v$ is of the order $v\,m_Q$. The scale $Q_v$
depends on the quark velocity $v$ and becomes soft for
$v\rightarrow0$, yielding the correct physical behavior. The PMC
scales in the usual $\overline{\rm MS}$ scheme are different from
the PMC scales in the physically-based V-scheme. This difference
is caused by the convention used in defining the $\overline{\rm
MS}$ scheme.

\begin{figure}[htb]
\centering
\includegraphics[width=0.40\textwidth]{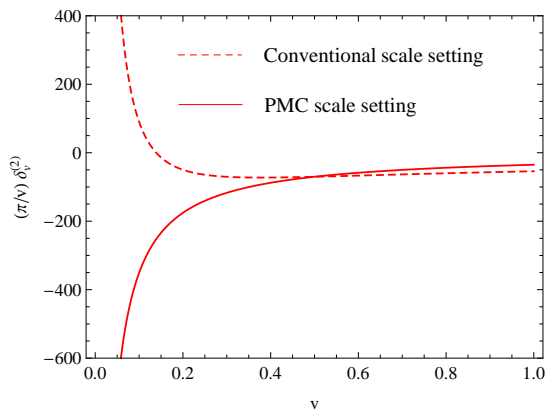}
\caption{The behavior of the Coulomb terms in the V-scheme for the
$b$ quark pair production, where
$\delta^{(2)}_v=(\delta^{(2)}_{v,in}|_V+\delta^{(2)}_{v,n_f}|_V\,n_f)$
for conventional scale setting, and for PMC scale setting,
$\delta^{(2)}_v=\delta^{(2)}_{v,\rm sc}|_V$. }
\label{figCoedetabV2}
\end{figure}

For the Coulomb correction, the behavior of the Coulomb term of
the form $(\pi/v)\,\delta^{(2)}_{v}$ is dramatically changed after
using the PMC. More explicitly, by taking $m_Q=4.89$ GeV for the
$b$ quark pair production as an example, we present the behavior
of the Coulomb terms of the form $(\pi/v)\,\delta^{(2)}_{v}$ in
the V-scheme using the conventional and the PMC scale setting in
Figure~\ref{figCoedetabV2}. Using the conventional scale setting in
the region where the quark velocity $v\rightarrow0$, the Coulomb
term becomes $(\pi/v)\delta^{(2)}_v\rightarrow+\infty$. On the
contrary, applying PMC scale setting, the Coulomb term becomes
$(\pi/v)\delta^{(2)}_v\rightarrow-\infty$. This dramatically
different behavior of the $(\pi/v)\delta^{(2)}_v$ between
conventional and PMC scale settings near the threshold region
should also be investigated in QED.

In analogy to the quark pair production, the lepton pair
production cross-section for the QED process
$e^{+}e^{-}\rightarrow\gamma^{*}\rightarrow l\bar{l}$ has an
expansion in the QED fine structure constant $\alpha_e$. The cross-section can also be divided into the non-Coulomb and Coulomb
parts, as in Equation~(\ref{eqpairNNLO}). The perturbative
coefficients for the lepton pair production cross-section are
given in Refs.~\cite{Czarnecki:1997vz, Hoang:1995ex, Hoang:1997sj}.

The one-loop correction coefficients $\delta^{(1)}_h$ and
$\delta^{(1)}_{v}$ and the two-loop correction coefficients
$\delta^{(2)}_{h,n_f}$, $\delta^{(2)}_{v,n_f}$, and
$\delta^{(2)}_{v^2}$ have the same form in QCD and QED, with only
some replacements for the color factors, i.e., $C_A=3$, $C_F=4/3$
and $T_R=1/2$ for QCD and $C_A=0$, $C_F=1$, and $T_R=1$ for QED,
respectively.

By using the PMC, the vacuum polarization corrections can be
absorbed into the QED running coupling whose one-loop
approximation is given by:
\begin{eqnarray}
\alpha_e(Q) = \alpha_e\left[1 + \left({\alpha_e\over\pi}\right)\sum^{n_f}\limits_{i=1}{1\over3}\left(\ln\left({Q^2\over m^2_i}\right)-{5\over3}\right)\right],
\end{eqnarray}
where $m_i$ is the mass of the light virtual lepton. The resulting PMC
scales can be written as
\begin{eqnarray}
Q_i = m_l\,\exp\left[{5\over6}+{3\over2}\,{\delta^{(2)}_{i,n_f}\over\delta^{(1)}_i}\right].
\end{eqnarray}

For the lepton pair production, we
obtain the PMC scales $Q_h =e^{(3/8)}\,m_l$ for the non-Coulomb
correction and $Q_v =2\,v\,m_l$ for the Coulomb correction.

Given that the scale $Q_h$ stems from the hard virtual photon
corrections, while $Q_v$ originates from the Coulomb rescattering,
it follows that $Q_h$ is of order $m_l$ and $Q_v$ is of order
$v\,m_l$. The scales show the same physical behavior from QCD to
QED after using the PMC. The PMC scales in QCD with the V-scheme
coincide with the PMC scales in QED. This scale self-consistency
shows that the PMC procedure in QCD agrees with the standard
{Gell-Mann}-Low method~\cite{GellMann:1954fq} in QED for the quark
pair production.

\begin{figure}[htb]
\centering
\includegraphics[width=0.40\textwidth]{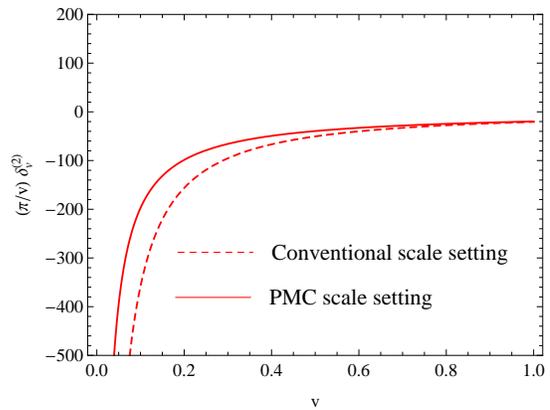}
\caption{The behavior of the Coulomb terms for the $\tau$ lepton
pair production, where
$\delta^{(2)}_v=(\delta^{(2)}_{v,in}+\delta^{(2)}_{v,n_f}\,n_f)$
for conventional scale setting, and for PMC scale setting,
$\delta^{(2)}_v=\delta^{(2)}_{v,\rm in}$. } \label{figCoeQEDb2}
\end{figure}

For the Coulomb correction, by taking $m_\tau=1.777$ GeV for the
$\tau$ lepton as an example, the behavior of the Coulomb terms of
the form $(\pi/v)\,\delta^{(2)}_{v}$ using conventional and PMC
scale settings is shown in Figure~\ref{figCoeQEDb2}. It is noted
that, different from the QCD case, when the quark velocity
$v\rightarrow0$, the Coulomb terms are
$(\pi/v)\delta^{(2)}_v\rightarrow-\infty$ for both the
conventional and the PMC scale settings in lepton pair production.
Thus, the behavior of the Coulomb terms is the same when using PMC
scale setting for both QCD and QED.

\subsection{Reanalysis of the Top-Quark Decay at Next-To-Next-To-Leading Order}

Top-quark properties, such as its mass, its production
cross-section, its decay width, and its couplings to elementary
particles, are very important for understanding the mechanism of
electro-weak symmetry breaking and for searching new physics
beyond the SM. At present, the top-quark decay width has been
calculated up to NNLO QCD \mbox{corrections~\cite{Czarnecki:1998qc,
Chetyrkin:1999ju, Blokland:2004ye, Blokland:2005vq, Gao:2012ja,
Brucherseifer:2013iv}}. Experimentally, various collaborations at
the Tevatron and LHC have measured the total width of the
top-quark decay, and the world average reported by the Particle
Data Group is $\Gamma_t=1.42^{+0.19}_{-0.15}$ GeV~\cite{PDG:2020}.

The top-quark decay process is dominated by $t\rightarrow bW$,
and its total decay width up to NNLO QCD corrections is given by:
\begin{eqnarray}
\Gamma_t=\Gamma^{\rm LO}_t\left[1+c_1(\mu_r) \, a_s(\mu_r)+c_2(\mu_r) \, a^2_s(\mu_r)+{\cal O}(\alpha^3_s)\right],
\label{Gamma_NNLO}
\end{eqnarray}
where the LO decay width
\begin{eqnarray}
\Gamma^{\rm LO}_t=\frac{G_F\,|V_{tb}|^2\,m_t^3}{8\,\pi\,\sqrt{2}}\left(1-3\,w^2+2\,w^3\right).
\end{eqnarray}

Here, $w={m^2_W}/{m^2_t}$, with $m_W=80.385$ GeV being the $W$-boson
mass, \mbox{$m_t=172.5$ GeV~\cite{PDG:2020}} is the top-quark mass,
$G_F=1.16638\times10^{-5}$ GeV$^{-2}$~\cite{Gao:2012ja} is the
Fermi constant, and $|V_{tb}|=1$ is the Cabibbo--Kobayashi--Maskawa
(CKM) matrix element. The NLO and NNLO coefficients $c_1$ and
$c_2$ can be found in the literature, and a detailed
PMC analysis for the top-quark decay process can be found in
Ref.~\cite{Meng:2022htg}. For self-consistency, we will adopt the
two-loop $\overline{\rm MS}$ QCD coupling with
$\alpha_s(M_Z)=0.1179$~\cite{PDG:2020} for numerical analysis.

We present the total decay width of the top-quark decay using the
conventional and PMC scale settings in Table \ref{tab1}, where the
NLO and NNLO contributions $\delta\Gamma^{\rm NLO}_t$ and
$\delta\Gamma^{\rm NNLO}_t$ are also shown. Up to the NNLO level, the
net scale uncertainty is $\sim[-0.5\%,+0.4\%]$ by varying the
scale $\mu_r$ within the range $[m_t/2,2m_t]$. Such a small net
scale uncertainty is due to the cancellation of the scale
uncertainties between $\delta\Gamma^{\rm NLO}_t$ and
$\delta\Gamma^{\rm NNLO}_t$. However, the scale uncertainty is
still rather large for each perturbative term, i.e., the scale
uncertainties of $\delta\Gamma^{\rm NLO}_t$ and
$\delta\Gamma^{\rm NNLO}_t$ are $\sim[-10.5\%,+7.9\%]$ and
$\sim[+23.5\%, -16.7\%]$, respectively.

If we set $\mu_r=m_t$, the relative importance of the NLO and NNLO
QCD corrections become $\delta\Gamma^{\rm NLO}_t/\Gamma^{\rm
LO}_t\sim-8.6\%$ and $\delta\Gamma^{\rm NNLO}_t/\Gamma^{\rm
LO}_t\sim-2.1\%$, respectively. The relative importance of each
order of accuracy up to NNLO becomes: $\delta\Gamma^{\rm
NLO}_t/\Gamma^{\rm LO}_t\sim-9.4\%$ and $\delta\Gamma^{\rm
NNLO}_t/\Gamma^{\rm LO}_t\sim-1.6\%$ for $\mu_r=m_t/2$; and
$\delta\Gamma^{\rm NLO}_t/\Gamma^{\rm LO}_t\sim-7.8\%$ and
$\delta\Gamma^{\rm NNLO}_t/\Gamma^{\rm LO}_t \sim-2.4\%$ for
$\mu_r=2m_t$. Thus, by using the conventional scale-setting
method, one cannot judge {a posteriori} the intrinsic
convergence of the pQCD series; a poorer convergent behavior may
be caused by improper choice of renormalization scale. This
explains why the renormalization scale uncertainty is one of the
systematic errors for pQCD predictions.

\begin{table} [htb]
\begin{tabular}{|c||c|c|c|c|c|}
\hline
~~ ~~  & ~scale $\mu_r$~ & ~$\Gamma^{\rm LO}_t$~ & ~$\delta\Gamma^{\rm NLO}_t$~ & ~$\delta\Gamma^{\rm NNLO}_t$~ & ~$\Gamma^{\rm NNLO}_t$~ \\
\hline
~~ ~~  & ~$\mu_r=m_t/2$~ & ~1.4806~ & ~-0.1394~ & ~-0.0234~ & ~1.3179~ \\
~Conv.~& ~$\mu_r=m_t$~   & ~1.4806~ & ~-0.1261~ & ~-0.0306~ & ~1.3239~  \\
~~ ~~  & ~$\mu_r=2m_t$~  & ~1.4806~ & ~-0.1161~ & ~-0.0357~ & ~1.3288~  \\
\hline
~PMC~ & ~~         & ~1.4806~ & ~-0.1892~ & ~0.0207~  & ~1.3122~ \\
\hline
\end{tabular}
\caption{The total decay width $\Gamma_t$ up to NNLO QCD corrections (in unit GeV) using the conventional (Conv.) and PMC scale settings, respectively.
\label{tab1} }
\end{table}

On the other hand, after applying the PMC scale setting, Table
\ref{tab1} shows that the PMC predictions are scale invariant, e.g.,
$\delta\Gamma^{\rm NLO}_t=-0.1892$ GeV and $\delta\Gamma^{\rm
NNLO}_t=0.0207$ GeV for any choice of $\mu_r$. This leads to a
scale invariant relative importance of the NLO and NNLO QCD
corrections, e.g., $\delta\Gamma^{\rm NLO}_t/\Gamma^{\rm
LO}_t\sim-12.8\%$ and $\delta\Gamma^{\rm NNLO}_t/\Gamma^{\rm
LO}_t\sim1.4\%$ for any choice of scale. Thus, with respect to the
conventional pQCD series for the top-quark decay, the convergence
of the PMC series is greatly improved.

The determined PMC scale for the top-quark decay is
\begin{eqnarray}
Q=15.5 ~\rm {GeV}.
\end{eqnarray}

The PMC scale is independent of any choice of $\mu_r$ and is one
order of magnitude smaller than $m_t$. This reflects the small
virtuality of the QCD dynamics for the top-quark decay process.
Numerically, we observe that the top-quark decay width at NNLO
first decreases and then increases with the increase in the
scale $\mu_r$ using a conventional scale setting, and the minimum
total decay width is achieved at $\mu_r\sim 23$ GeV. If we change
the conventional choice $\mu_r=m_t$ to a smaller-scale $\mu_r\ll
m_t$, the pQCD convergence of the top-quark decay width would be
greatly improved, even though the resulting conventional
prediction is close to the PMC prediction. Thus, the effective
momentum flow for the top-quark decay should be $\ll m_t$, far
smaller than the conventional choice of $\mu_r=m_t$.

After applying the PMC in order to achieve reliable predictions,
there are still other error sources that have to be taken into
account, such as the effect of finite bottom-quark mass and the
finite $W$ boson width, as well as the electroweak corrections. In
Table~\ref{tab3}, we present the top-quark decay width $\Gamma^{\rm
NNLO}_t|_{\rm PMC}$ using the PMC together with the corrections
from the finite bottom-quark mass $\delta_f^b$, the finite $W$
boson width $\delta_f^W$, and the NLO electroweak correction
$\delta^{\rm NLO}_{\rm EW}$ for $m_t=172.5$ and $173.5$ GeV. These
corrections are taken from Ref.~\cite{Gao:2012ja}. Since the
corrections from the finite bottom-quark mass and the finite $W$
boson width provide negative values, while the NLO electroweak
correction provides a positive value, their contributions to the
top-quark decay width cancel out greatly.

\begin{table} [htb]
\begin{tabular}{|c|c|c|c|c|c|}
\hline
 $m_t$        & $\Gamma^{\rm NNLO}_t|_{\rm PMC}$  & $\delta_f^b$ & $\delta_f^W$  & $\delta^{\rm NLO}_{\rm EW}$ & $\Gamma^{\rm tot}_t$ \\
\hline
~172.5~       & ~$1.3122$~ & ~-0.0038~ & ~-0.0221~  & ~0.0249~ & ~1.3112~   \\
\hline
~173.5~       & ~$1.3392$~ & ~-0.0039~ & ~-0.0225~  & ~0.0255~ & ~1.3383~   \\
\hline
\end{tabular}
\caption{The PMC top-quark decay widths $\Gamma^{\rm NNLO}_t|_{\rm
PMC}$ (in unit GeV). Uncertainties caused by the bottom quark mass
$\delta_f^b$, the finite $W$-boson width $\delta_f^W$ and the NLO
electroweak corrections $\delta^{\rm NLO}_{\rm EW}$ are also
presented, whose magnitudes are taken from Ref.\cite{Gao:2012ja}.
\label{tab3} }
\end{table}

After applying the PMC, we then obtain more reliable predictions
for the top-quark total decay width~\cite{Meng:2022htg}. If we set
$m_t=172.5$ GeV, we have
\begin{eqnarray}
\Gamma^{\rm tot}_t&=&1.3112\pm 0.0016\pm 0.0023 ~\rm {GeV},
\end{eqnarray}
whhereas if we set $m_t=173.5$ GeV, we have
\begin{eqnarray}
\Gamma^{\rm tot}_t&=&1.3383^{+0.0016}_{-0.0017}\pm 0.0023 ~\rm {GeV}.
\end{eqnarray}

\begin{figure}[htb]
\centering
\includegraphics[width=0.40\textwidth]{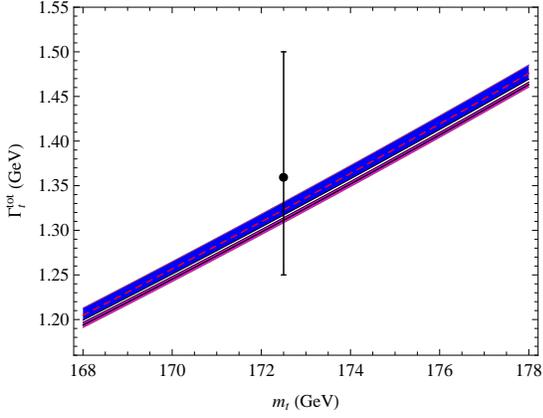}
\caption{The top-quark total decay width $\Gamma^{\rm tot}_t$
versus the top-quark mass $m_t$. The solid line is the PMC
prediction, and the dashed line stands for the conventional
prediction. As a comparison, the CMS measured
value~\cite{CMS:2014mxl} is also presented. } \label{topdecaymt}
\end{figure}

The first error is due to the error of  $\alpha_s$ at the critical
scale $\mu_r=M_Z$, e.g., \linebreak $\Delta
\alpha_s(M_Z)=\pm0.0009$~\cite{PDG:2020}, and the second error is
given by the evaluation of the UHO terms. The
top-quark total decay width depends heavily on the magnitude of
the top-quark mass. More explicitly, we present the top-quark
total decay width $\Gamma^{\rm tot}_t$ versus $m_t$ in
Figure~\ref{topdecaymt}. The CMS measurement together with its
error~\cite{CMS:2014mxl} are also presented in
Figure~\ref{topdecaymt}. No discrepancy is observed in comparing the
experimental value and the theoretical predictions obtained by
using the PMC and the conventional scale setting, respectively
(Figure~\ref{topdecaymt}).

\subsection{An Estimate of the Contributions from Uncalculated Higher-Order Terms}

At present, remarkable progress has been achieved in loop calculations in perturbation theory. However, most of theoretical calculations have only been finished at relatively lower orders due to the complexity of Feynman diagram calculations. It is thus important to have a way to evaluate contributions from the UHO terms, in order to improve the predictive power of perturbative theory.

In this subsection, we briefly review two representative
approaches to evaluate the UHO contributions for the fixed-order pQCD series by using the known partial sum of the conventional series and PMC conformal series, respectively. The first approach is the Pad$\acute{\rm e}$ approximation approach
(PAA)~\cite{Basdevant:1972fe, Samuel:1992qg, Samuel:1995jc}, which attempts to directly predict the unknown higher-order coefficients by using a fractional generating function whose parameters can be directly fixed by matching to the known finite order. The
second approach is the Bayesian-based approach (BA) \cite{Cacciari:2011ze, Bagnaschi:2014wea, Bonvini:2020xeo, Duhr:2021mfd}, which attempts to quantify the unknown higher-order terms in terms of
a probability distribution by applying Bayes' theorem.

\subsubsection{Applying PAA to Evaluate the UHO Contributions}

The Pad$\acute{\rm e}$ approximation provides a feasible approach that predicts the unknown $(n+1)_{\rm th}$-order coefficient from the known $n_{\rm th}$-order perturbative series. Following the basic PAA procedure, an $[N/M]$-type fractional generating function $\rho_n^{[N/M]}$ for $\rho_n=\sum_{i=0}^{n(\geq 1)}c_i \alpha_s^i$ is constructed as~\cite{Basdevant:1972fe, Samuel:1992qg, Samuel:1995jc}
\begin{eqnarray}
\rho^{[N/M]}_n &=& \frac{d_0+d_1 \alpha_s + \cdots + d_N \alpha_s^N} {1 + e_1 \alpha_s + \cdots + e_M \alpha_s^M} \nonumber \\
               &=& \sum_{i=0}^{n} c_i \alpha_s^i + c_{n+1} \alpha_s^{n+1} +\cdots,
\end{eqnarray}
where $M\geq 1$ and $N+M=n$. The parameters $d_{i}\;(0\leq i\leq N)$ and $e_{j}\;(1\leq j\leq M)$ are firstly determined by the known coefficients $c_{i}\;(0\leq i\leq n)$ and then provide a reasonable prediction for the next uncalculated coefficient $c_{n+1}$. For $n=4$, it has been found that the diagonal [2/2]-type generating function is preferable for predicting unknown coefficients from the conventional pQCD series \cite{Gardi:1996iq, Cvetic:1997qm}, while the non-diagonal [0/4]-type generating function is preferable for predicting unknown coefficients from the PMC conformal series~~\cite{Du:2018dma}, which also expands the geometric series to be self-consistent with the GM-L prediction \cite{GellMann:1954fq}.

\subsubsection{Applying BA to Evaluate the UHO Contributions}

The BA quantifies the UHO coefficients in terms of probability distributions, in which Bayes' theorem is applied to iteratively update the probability as new coefficients become available. Here we present the main results for BA; for a detailed introduction and all BA formulas, see, e.g., \cite{Shen:2022nyr} and references therein.

Following the BA procedure, the conditional probability density function (p.d.f.) for a generic (uncalculated) coefficient $c_{n}$ ($n>k$) of any possible perturbative series $\rho_k=\sum_{i=1}^{k}c_i\alpha_s^i$ with given coefficients $\{c_1,c_2,\dots,c_k\}$ is given by
\begin{eqnarray}\label{eq:cnpdf}
\hspace{-3mm}
f(c_{n}|c_1,c_2,\dots,c_k) = \left\{ \begin{array}{l l}
\frac{k}{2(k+1)\bar{c}_{(k)}}, &    |c_n| \leq \bar{c}_{(k)} \vspace{1mm}\\
\frac{k{\bar{c}_{(k)}^{k}}}{2(k+1)|c_n|^{k+1}}, & |c_n| > \bar{c}_{(k)} \\
\end{array}
\right.,
\end{eqnarray}
where $\bar{c}_{(k)}=\max\{|c_1|,|c_2|,\cdots,|c_k|\}$.
Equation~(\ref{eq:cnpdf}) provides a symmetric probability distribution
for negative and positive $c_n$, predicts a uniform probability
density in the interval $[-\bar{c}_{(k)},\bar{c}_{(k)}]$, and
decreases monotonically from $\bar{c}_{(k)}$ to infinity. The
knowledge of probability density $f_c(c_{n}|c_1,c_2,\dots,c_k)$
allows one to calculate the degree-of-belief (DoB) that the value
of $c_{n}$ belongs to some credible interval (CI). The symmetric
smallest CI of fixed $p\%$ DoB for $c_{n}$ is denoted by
$[-c_n^{(p)},c_n^{(p)}]$. Here the boundary $c_n^{(p)}$ is defined
implicitly by $p\% = \int_{-c_n^{(p)}}^{c_n^{(p)}} f_c(c_n
|c_l,\dots,c_k)\ {\rm d} c_n$ and can be obtained by further
using the analytical expression in Equation~(\ref{eq:cnpdf}),
\begin{eqnarray}
\label{eq:cnp}
\hspace{-3mm}
c_n^{(p)}=
\left\{
\begin{array}{l l}
\bar{c}_{(k)} \frac{k+1}{k} p\%, & \;  p\% \leq \frac{k}{k+1} \vspace{2mm}\\
\bar{c}_{(k)} \left[(k+1)(1-p\%)\right]^{-\frac{1}{k}}, & \; p\% > \frac{k}{k+1}   \\
\end{array}
\right..
\end{eqnarray}

We adopt the interval
$[-c_n^{(p)}\alpha_s^{n},c_n^{(p)}\alpha_s^{n}]$ with $p\%=95.5\%$\footnote{One may also use a $68.3\%$ credible interval (CI) to
compare with experimental data in the same confidence level, or
use $99.7\%$ CI for a more conservative estimation.} as the final
estimation for any UHO term $\delta_n=c_n\alpha_s^n$.

\begin{figure}[htb]
\centering
\includegraphics[width=0.40\textwidth]{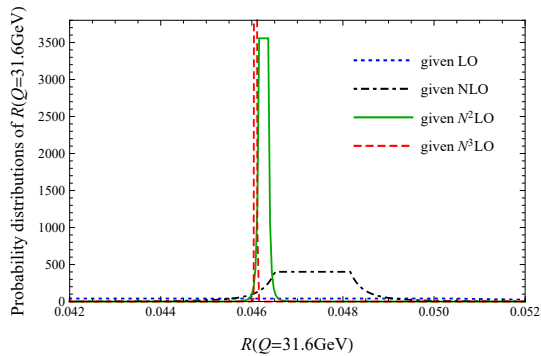}
\caption{The probability density distributions of $R(Q=31.6\;{\rm GeV})$ with different states of knowledge predicted by PMCs and BA. The
blue dotted, the black dash-dotted, the green solid, and the red
dashed curves represent the results with given LO, NLO, N$^2$LO, and
N$^3$LO series, respectively.}\label{fig:Reedistribution}
\end{figure}

As an example, we consider the total hadronic $e^+e^-$ annihilation ratio $R_{e^+ e^-}(Q) = N_c\sum_q e_q^2\left[1+R(Q)\right]$,
where $R(Q)$ represents the QCD correction. The probability density distributions for $R(Q=31.6\;{\rm GeV})$ with different states of knowledge predicted by PMCs and BA are presented in
Figure~\ref{fig:Reedistribution}, where the four curves correspond to different degrees of knowledge: given LO (dotted),
given NLO (dot-dashed), given N$^2$LO (solid), and given N$^3$LO
(dashed). The figure illustrates the characteristics of the
posterior probability distribution: a symmetric plateau with two suppressed tails. The posterior probability distribution depends on the prior probability distribution. With more and more loop terms
available, the posterior probability is updated and becomes less and less dependent on the prior probability; i.e., the probability density becomes increasingly concentrated as more and more loop terms are added.

As a final remark, the PAA and BA can only be applied after one has specified the choice for the renormalization scale due to the fact that the coefficients of the conventional pQCD series are scale-dependent. Thus, extra uncertainties are introduced when applying the PAA and BA to a conventional pQCD series. However, the resulting PMC series is scale-independent, and it thus provides a more
reliable basis for estimating the UHO contributions. Thus, the total theoretical uncertainty of a pQCD approximant can be treated as the squared average of the scale error due to the conventional scale dependence (or the first kind of residual scale dependence) and the predicted magnitude of the UHO terms for the pQCD approximant.

\section{Summary}

The PMC method provides a systematic way to eliminate the renormalization scheme-and-scale ambiguities.
The PMC method has a rigorous theoretical foundation, satisfying the
RGI and all of the self-consistency conditions derived from the
renormalization group. The PMC scales are obtained by shifting the
argument of $\alpha_s$ to eliminate all the non-conformal
$\beta$-terms; the PMC scales thus reflect the virtuality of the
propagating gluons for the QCD processes. The divergent renormalon
contributions are eliminated since they are summed into the
running coupling $\alpha_s$, and the resulting pQCD convergence is in
general greatly improved. The PMC scale-setting method provides
the underlying principle for the well-known BLM method, extending
the BLM scale-setting procedure unambiguously to all orders. The
PMC reduces to the GM-L method in the $N_C\to 0$ Abelian
limit~\cite{Brodsky:1997jk}.

We have provided a new analysis of event shape observables in $e^+e^-$
annihilation by using the PMC method. The PMC scales are not given by
a single value but depend dynamically on
the virtuality of the underlying quark and gluon subprocess and
thus the specific kinematics of each event. The renormalization scale-independent
PMC predictions for event shape distributions agree with precise
experimental data. Remarkably, the PMC method provides a novel
method for the precise determination of the running of
QCD coupling $\alpha_s(Q^2)$ over a wide range of $Q^2$ from event shapes
measured at a single energy of $\sqrt{s}$. The PMC
also provides an unambiguous method for determining the scales in
multiple-scale processes. It is remarkable that two distinctly
different PMC scales are determined for the heavy fermion pair
production near the threshold region. One PMC scale entering the hard
virtual corrections is of the order of the fermion mass $m_f$, while
the other PMC scale entering the Coulomb re-scattering amplitude is
of the order $ v\,m_f$. Perfect agreement between the Abelian
unambiguous {Gell-Mann}-Low and the PMC scale-setting method in the
limit of zero number of colors is observed in this process. We also
calculated the top-quark decay process, obtaining the PMC scale
$Q=15.5$ GeV. The convergence of the pQCD series
is largely improved for the top-quark decay. We finally obtained
the top-quark total decay width $\Gamma^{\rm
tot}_t=1.3112^{+0.0190}_{-0.0189}$ GeV. Since the PMC conformal
series is scale-independent, it provides a reliable basis for
obtaining constraints on the predictions for the UHO
contributions. These applications demonstrate the generality and
applicability of the PMC. The PMC thus improves the precision
tests of the SM and and increases the sensitivity of experiments
to new physics beyond the SM.

\hspace{1cm}

{\bf Acknowledgements}: This work was supported in part by the
Natural Science Foundation of China under Grants No.12265011,
No.12175025 and No.12147102; by the Project of Guizhou Provincial
Department under Grant No.KY[2021]003, No.GZMUZK[2022]PT01 and No.ZK[2023]141; and by
the Department of Energy (DOE), Contract DECAC02C76SF00515.
SLAC-PUB-17723.


\begin{thebibliography}{99}

\bibitem{Politzer:1973fx}
  H.~D.~Politzer,
  Reliable Perturbative Results for Strong Interactions ?,
  Phys. Rev. Lett. \textbf{30}, 1346 (1973).

\bibitem{Gross:1973id}
  D.~J.~Gross and F.~Wilczek,
  Ultraviolet Behavior of Nonabelian Gauge Theories,
  Phys. Rev. Lett. \textbf{30}, 1343 (1973).

\bibitem{Beneke:1998ui}
  M.~Beneke,
  Renormalons,
  Phys.\ Rept.\  {\bf 317}, 1 (1999).

\bibitem{GellMann:1954fq}
  M.~Gell-Mann and F.~E.~Low,
  Quantum electrodynamics at small distances,
  Phys.\ Rev.\ {\bf 95}, 1300 (1954).

\bibitem{Brodsky:1997jk}
 S.~J.~Brodsky and P.~Huet,
 Aspects of SU(N(c)) gauge theories in the limit of small number of colors,
 Phys.\ Lett.\ B {\bf 417}, 145 (1998).

\bibitem{Brodsky:1982gc}
  S.~J.~Brodsky, G.~P.~Lepage and P.~B.~Mackenzie,
  On the Elimination of Scale Ambiguities in Perturbative Quantum Chromodynamics,
  Phys.\ Rev.\ D {\bf 28}, 228 (1983).

\bibitem{Brodsky:2011ta}
  S.~J.~Brodsky and X.~G.~Wu,
  Scale Setting Using the Extended Renormalization Group and the Principle of Maximum Conformality: the QCD Coupling Constant at Four Loops,
  Phys.\ Rev.\ D {\bf 85}, 034038 (2012).

\bibitem{Brodsky:2012rj}
  S.~J.~Brodsky and X.~G.~Wu,
  Eliminating the Renormalization Scale Ambiguity for Top-Pair Production Using the Principle of Maximum Conformality,
  Phys.\ Rev.\ Lett.\  {\bf 109}, 042002 (2012).

\bibitem{Brodsky:2011ig}
  S.~J.~Brodsky and L.~Di Giustino,
  Setting the Renormalization Scale in QCD: The Principle of Maximum Conformality,
  Phys.\ Rev.\ D {\bf 86}, 085026 (2012).

\bibitem{Mojaza:2012mf}
  M.~Mojaza, S.~J.~Brodsky and X.~G.~Wu,
  Systematic All-Orders Method to Eliminate Renormalization-Scale and Scheme Ambiguities in Perturbative QCD,
  Phys.\ Rev.\ Lett.\  {\bf 110}, 192001 (2013).

\bibitem{Brodsky:2013vpa}
  S.~J.~Brodsky, M.~Mojaza and X.~G.~Wu,
  Systematic Scale-Setting to All Orders: The Principle of Maximum Conformality and Commensurate Scale Relations,
  Phys.\ Rev.\ D {\bf 89}, 014027 (2014).

\bibitem{Wu:2013ei}
 X.~G.~Wu, S.~J.~Brodsky and M.~Mojaza,
  The Renormalization Scale-Setting Problem in QCD,
  Prog.\ Part.\ Nucl.\ Phys.\ {\bf 72}, 44 (2013).

\bibitem{Wu:2014iba}
  X.~G.~Wu, Y.~Ma, S.~Q.~Wang, H.~B.~Fu, H.~H.~Ma, S.~J.~Brodsky and M.~Mojaza,
  Renormalization Group Invariance and Optimal QCD Renormalization Scale-Setting,
  Rept.\ Prog.\ Phys.\  {\bf 78}, 126201 (2015).

\bibitem{Wu:2019mky}
  X.~G.~Wu, J.~M.~Shen, B.~L.~Du, X.~D.~Huang, S.~Q.~Wang and S.~J.~Brodsky,
  The QCD renormalization group equation and the elimination of fixed-order scheme-and-scale ambiguities using the principle of maximum conformality,
  Prog.\ Part.\ Nucl.\ Phys.\  {\bf 108}, 103706 (2019).

\bibitem{Brodsky:2012ms}
  S.~J.~Brodsky and X.~G.~Wu,
  Self-Consistency Requirements of the Renormalization Group for Setting the Renormalization Scale,
  Phys.\ Rev.\ D {\bf 86}, 054018 (2012).

\bibitem{Shen:2017pdu}
  J.~M.~Shen, X.~G.~Wu, B.~L.~Du and S.~J.~Brodsky,
  Novel All-Orders Single-Scale Approach to QCD Renormalization Scale-Setting,
  Phys. Rev. D \textbf{95}, 094006 (2017).

\bibitem{DiGiustino:2020fbk}
  L.~Di Giustino, S.~J.~Brodsky, S.~Q.~Wang and X.~G.~Wu,
  Infinite-order scale-setting using the principle of maximum conformality: A remarkably efficient method for eliminating renormalization scale ambiguities for perturbative QCD,
  Phys. Rev. D {\bf 102}, 014015 (2020).

\bibitem{Yan:2022foz}
  J.~Yan, Z.~F.~Wu, J.~M.~Shen and X.~G.~Wu,
  Precise perturbative predictions from fixed-order calculations,
  J. Phys. G \textbf{50}, 045001 (2023).

\bibitem{Bi:2015wea}
  H.~Y.~Bi, X.~G.~Wu, Y.~Ma, H.~H.~Ma, S.~J.~Brodsky and M.~Mojaza,
  Degeneracy Relations in QCD and the Equivalence of Two Systematic All-Orders Methods for Setting the Renormalization Scale,
  Phys. Lett. B \textbf{748}, 13 (2015).

\bibitem{Zheng:2013uja}
  X.~C.~Zheng, X.~G.~Wu, S.~Q.~Wang, J.~M.~Shen and Q.~L.~Zhang,
  Reanalysis of the BFKL Pomeron at the next-to-leading logarithmic accuracy,
  JHEP {\bf 1310}, 117 (2013).

\bibitem{Deur:2017cvd}
  A.~Deur, J.~M.~Shen, X.~G.~Wu, S.~J.~Brodsky and G.~F.~de Teramond,
  Implications of the Principle of Maximum Conformality for the QCD Strong Coupling,
  Phys.\ Lett.\ B {\bf 773}, 98 (2017).

\bibitem{Yu:2021yvw}
  Q.~Yu, H.~Zhou, X.~D.~Huang, J.~M.~Shen and X.~G.~Wu,
  Novel and self-consistency analysis of the QCD running coupling $\alpha_s(Q)$ in both the perturbative and nonperturbative domains,
  Chin. Phys. Lett. {\bf 39}, 071201 (2022).

\bibitem{Wu:2015rga}
  X.~G.~Wu, S.~Q.~Wang and S.~J.~Brodsky,
  Importance of proper renormalization scale-setting for QCD testing at colliders,
  Front. Phys. (Beijing) \textbf{11}, 111201 (2016).

\bibitem{Meng:2022htg}
  R.~Q.~Meng, S.~Q.~Wang, T.~Sun, C.~Q.~Luo, J.~M.~Shen and X.~G.~Wu,
  QCD improved top-quark decay at next-to-next-to-leading order,
  Eur.\ Phys.\ J.\ C {\bf 83}, 59 (2023).

\bibitem{Brodsky:1999je}
 S.~J.~Brodsky, V.~S.~Fadin, V.~T.~Kim, L.~N.~Lipatov, and G.~B.~Pivovarov,
  The QCD pomeron with optimal renormalization,
  JETP Lett. {\bf 70}, 155 (1999).

\bibitem{Hentschinski:2012kr}
  M.~Hentschinski, A.~Sabio Vera and C.~Salas,
  Hard to Soft Pomeron Transition in Small-x Deep Inelastic Scattering Data Using Optimal Renormalization,
  Phys.\ Rev.\ Lett.\  {\bf 110}, 041601 (2013).

\bibitem{Caporale:2015uva}
  F.~Caporale, D.~Y.~Ivanov, B.~Murdaca and A.~Papa,
  Brodsky-Lepage-Mackenzie optimal renormalization scale setting for semihard processes,
  Phys.\ Rev.\ D {\bf 91}, 114009 (2015).

\bibitem{Wang:2021tak}
  S.~Q.~Wang, C.~Q.~Luo, X.~G.~Wu, J.~M.~Shen and L.~Di Giustino,
  New analyses of event shape observables in electron-positron annihilation and the determination of \ensuremath{\alpha}$_{s}$ running behavior in perturbative domain,
  JHEP \textbf{09}, 137 (2022).

\bibitem{Wang:2014wua}
  S.~Q.~Wang, X.~G.~Wu, J.~M.~Shen, H.~Y.~Han and Y.~Ma,
  QCD improved electroweak parameter \ensuremath{\rho},
  Phys. Rev. D \textbf{89}, 116001 (2014).

\bibitem{Yu:2021ujk}
  Q.~Yu, H.~Zhou, J.~Yan, X.~D.~Huang and X.~G.~Wu,
  A new analysis of the pQCD contributions to the electroweak parameter \ensuremath{\rho} using the single-scale approach of principle of maximum conformality,
  Phys. Lett. B \textbf{820}, 136574 (2021).

\bibitem{Shen:2015cta}
  J.~M.~Shen, X.~G.~Wu, H.~H.~Ma, H.~Y.~Bi and S.~Q.~Wang,
  Renormalization group improved pQCD prediction for $\Upsilon(1S)$ leptonic decay,
  JHEP {\bf 1506}, 169 (2015).

\bibitem{Huang:2019frb}
  X.~D.~Huang, X.~G.~Wu, J.~Zeng, Q.~Yu and J.~M.~Shen,
  The $\Upsilon (1S)$ leptonic decay using the principle of maximum conformality,
  Eur.\ Phys.\ J.\ C {\bf 79}, 650 (2019).

\bibitem{Wang:2013vn}
  S.~Q.~Wang, X.~G.~Wu, X.~C.~Zheng, J.~M.~Shen and Q.~L.~Zhang,
  $J/\psi +\chi_{cJ}$ Production at the $B$ Factories under the Principle of Maximum Conformality,
  Nucl.\ Phys.\ B {\bf 876}, 731 (2013).

\bibitem{Sun:2018rgx}
  Z.~Sun, X.~G.~Wu, Y.~Ma and S.~J.~Brodsky,
  Exclusive production of $J/\psi+\eta_c$ at the $B$ factories Belle and Babar using the principle of maximum conformality,
  Phys.\ Rev.\ D {\bf 98}, 094001 (2018).

\bibitem{Yu:2020tri}
  H.~M.~Yu, W.~L.~Sang, X.~D.~Huang, J.~Zeng, X.~G.~Wu and S.~J.~Brodsky,
  Scale-fixed predictions for $\gamma + \eta_c$ production in electron-positron collisions at NNLO in perturbative QCD,
  JHEP \textbf{01}, 131 (2021).

\bibitem{Qiao:2014lwa}
  C.~F.~Qiao, R.~L.~Zhu, X.~G.~Wu and S.~J.~Brodsky,
  A possible solution to the B->PIPI puzzle using the principle of maximum conformality,
  Phys.\ Lett.\ B {\bf 748}, 422 (2015).

\bibitem{Wang:2018lry}
  S.~Q.~Wang, X.~G.~Wu, W.~L.~Sang and S.~J.~Brodsky,
  Solution to the $\gamma\gamma^*\rightarrow\eta_c$ puzzle using the principle of maximum conformality,
  Phys.\ Rev.\ D {\bf 97}, 094034 (2018).

\bibitem{PDG:2020}
  P.~A.~Zyla {\it et al.} (Particle Data Group), Prog. Theor. Exp. Phys. 2020, 083C01 (2020).

\bibitem{Heister:2003aj}
  A.~Heister {\it et al.} [ALEPH Collaboration],
  Studies of QCD at $e^+ e^-$ centre-of-mass energies between 91 GeV and 209 GeV,
  Eur.\ Phys.\ J.\ C {\bf 35}, 457 (2004).

\bibitem{Abdallah:2003xz}
  J.~Abdallah {\it et al.} [DELPHI Collaboration],
  A Study of the energy evolution of event shape distributions and their means with the DELPHI detector at LEP,
  Eur.\ Phys.\ J.\ C {\bf 29}, 285 (2003).

\bibitem{Abbiendi:2004qz}
  G.~Abbiendi {\it et al.} [OPAL Collaboration],
  Measurement of event shape distributions and moments in $e^{+} e^{-}\rightarrow$ hadrons at 91-209 GeV and a determination of $\alpha_s$,
  Eur.\ Phys.\ J.\ C {\bf 40}, 287 (2005).

\bibitem{Achard:2004sv}
  P.~Achard {\it et al.} [L3 Collaboration],
  Studies of hadronic event structure in $e^{+} e^{-}$ annihilation from 30 to 209 GeV with the L3 detector,
  Phys.\ Rept.\  {\bf 399}, 71 (2004).

\bibitem{Abe:1994mf}
  K.~Abe {\it et al.} [SLD Collaboration],
  Measurement of $\alpha_s(M_Z^2)$ from hadronic event observables at the Z$^0$ resonance,
  Phys.\ Rev.\ D {\bf 51}, 962 (1995).

\bibitem{Gehrmann-DeRidder:2007nzq}
  A.~Gehrmann-De Ridder, T.~Gehrmann, E.~W.~N.~Glover and G.~Heinrich,
  Second-order QCD corrections to the thrust distribution,
  Phys.\ Rev.\ Lett.\  {\bf 99}, 132002 (2007).

\bibitem{GehrmannDeRidder:2007hr}
  A.~Gehrmann-De Ridder, T.~Gehrmann, E.~W.~N.~Glover and G.~Heinrich,
  NNLO corrections to event shapes in $e^+ e^-$ annihilation,
  JHEP {\bf 0712}, 094 (2007).

\bibitem{Ridder:2014wza}
  A.~Gehrmann-De Ridder, T.~Gehrmann, E.~W.~N.~Glover and G.~Heinrich,
  EERAD3: Event shapes and jet rates in electron-positron annihilation at order $\alpha_s^3$,
  Comput.\ Phys.\ Commun.\  {\bf 185}, 3331 (2014).

\bibitem{Weinzierl:2008iv}
  S.~Weinzierl,
  NNLO corrections to 3-jet observables in electron-positron annihilation,
  Phys.\ Rev.\ Lett.\  {\bf 101}, 162001 (2008).

\bibitem{Weinzierl:2009ms}
  S.~Weinzierl,
  Event shapes and jet rates in electron-positron annihilation at NNLO,
  JHEP {\bf 0906}, 041 (2009).

\bibitem{DelDuca:2016csb}
  V.~Del Duca, C.~Duhr, A.~Kardos, G.~Somogyi and Z.~Tr\'{o}cs\'{a}nyi,
  Three-Jet Production in Electron-Positron Collisions at Next-to-Next-to-Leading Order Accuracy,
  Phys.\ Rev.\ Lett.\  {\bf 117}, 152004 (2016).

\bibitem{DelDuca:2016ily}
  V.~Del Duca, C.~Duhr, A.~Kardos, G.~Somogyi, Z.~Sz\H{o}r, Z.~Tr\'{o}cs\'{a}nyi and Z.~Tulip\'{a}nt,
  Jet production in the CoLoRFulNNLO method: event shapes in electron-positron collisions,
  Phys.\ Rev.\ D {\bf 94}, 074019 (2016).

\bibitem{Wang:2019isi}
  S.~Q.~Wang, S.~J.~Brodsky, X.~G.~Wu, J.~M.~Shen and L.~Di Giustino,
  Novel method for the precise determination of the QCD running coupling from event shape distributions in electron-positron annihilation,
  Phys.\ Rev.\ D {\bf 100}, 094010 (2019).

\bibitem{Brandt:1964sa}
  S.~Brandt, C.~Peyrou, R.~Sosnowski and A.~Wroblewski,
  The Principal axis of jets. An Attempt to analyze high-energy collisions as two-body processes,
  Phys.\ Lett.\  {\bf 12}, 57 (1964).

\bibitem{Farhi:1977sg}
  E.~Farhi,
  A QCD Test for Jets,
  Phys.\ Rev.\ Lett.\  {\bf 39}, 1587 (1977).

\bibitem{Parisi:1978eg}
  G.~Parisi,
  Super Inclusive Cross-Sections,
  Phys.\ Lett.\ B {\bf 74}, 65 (1978).

\bibitem{Donoghue:1979vi}
  J.~F.~Donoghue, F.~E.~Low and S.~Y.~Pi,
  Tensor Analysis of Hadronic Jets in Quantum Chromodynamics,
  Phys.\ Rev.\ D {\bf 20}, 2759 (1979).

\bibitem{GehrmannDeRidder:2009dp}
  A.~Gehrmann-De Ridder, T.~Gehrmann, E.~W.~N.~Glover and G.~Heinrich,
  NNLO moments of event shapes in e+e- annihilation,
  JHEP {\bf 0905}, 106 (2009).

\bibitem{Weinzierl:2009yz}
  S.~Weinzierl,
  Moments of event shapes in electron-positron annihilation at NNLO,
  Phys.\ Rev.\ D {\bf 80}, 094018 (2009).

\bibitem{Pahl:2007zz}
  C.~J.~Pahl,
  CERN-THESIS-2007-188;

\bibitem{Czarnecki:1997vz}
  A.~Czarnecki and K.~Melnikov,
  Two loop QCD corrections to the heavy quark pair production cross-section in $e^+e^-$ annihilation near the threshold,
  Phys.\ Rev.\ Lett.\  {\bf 80}, 2531 (1998).

\bibitem{Beneke:1997jm}
  M.~Beneke, A.~Signer and V.~A.~Smirnov,
  Two loop correction to the leptonic decay of quarkonium,
  Phys.\ Rev.\ Lett.\  {\bf 80}, 2535 (1998).

\bibitem{Bernreuther:2006vp}
  W.~Bernreuther, R.~Bonciani, T.~Gehrmann, R.~Heinesch, T.~Leineweber, P.~Mastrolia and E.~Remiddi,
  Two-Parton Contribution to the Heavy-Quark Forward-Backward Asymmetry in NNLO QCD,
  Nucl.\ Phys.\ B {\bf 750}, 83 (2006).

\bibitem{Brodsky:1995ds}
  S.~J.~Brodsky, A.~H.~Hoang, J.~H.~Kuhn and T.~Teubner,
  Angular distributions of massive quarks and leptons close to threshold,
  Phys.\ Lett.\ B {\bf 359}, 355 (1995).

\bibitem{Appelquist:1977tw}
  T.~Appelquist, M.~Dine and I.~J.~Muzinich,
  The Static Potential in Quantum Chromodynamics,
  Phys.\ Lett.\  {\bf 69B}, 231 (1977).

\bibitem{Fischler:1977yf}
  W.~Fischler,
  Quark - anti-Quark Potential in QCD,
  Nucl.\ Phys.\ B {\bf 129}, 157 (1977).

\bibitem{Peter:1996ig}
  M.~Peter,
  The Static quark - anti-quark potential in QCD to three loops,
  Phys.\ Rev.\ Lett.\  {\bf 78}, 602 (1997).

\bibitem{Schroder:1998vy}
  Y.~Schroder,
  The Static potential in QCD to two loops,
  Phys.\ Lett.\ B {\bf 447}, 321 (1999).

\bibitem{Smirnov:2008pn}
  A.~V.~Smirnov, V.~A.~Smirnov and M.~Steinhauser,
  Fermionic contributions to the three-loop static potential,
  Phys.\ Lett.\ B {\bf 668}, 293 (2008).

\bibitem{Smirnov:2009fh}
  A.~V.~Smirnov, V.~A.~Smirnov and M.~Steinhauser,
  Three-loop static potential
  Phys.\ Rev.\ Lett.\  {\bf 104}, 112002 (2010).

\bibitem{Anzai:2009tm}
  C.~Anzai, Y.~Kiyo and Y.~Sumino,
  Static QCD potential at three-loop order,
  Phys.\ Rev.\ Lett.\  {\bf 104}, 112003 (2010).

\bibitem{Kataev:2015yha}
A.~L.~Kataev and V.~S.~Molokoedov,
  Fourth-order QCD renormalization group quantities in the V scheme and the relation of the $\beta$ function to the Gell-Mann\textendash{}Low function in QED,
  Phys. Rev. D \textbf{92} (2015), 054008.

\bibitem{Kataev:2023sru}
  A.~L.~Kataev and V.~S.~Molokoedov,
  The generalized Crewther relation and V-scheme: analytic $O(\alpha^4_s)$ results in QCD and QED,
  [arXiv:2302.03443 [hep-ph]].

\bibitem{Wang:2020ckr}
  S.~Q.~Wang, S.~J.~Brodsky, X.~G.~Wu, L.~Di Giustino and J.~M.~Shen,
  Renormalization scale setting for heavy quark pair production in $e^+e^-$ annihilation near the threshold region,
  Phys. Rev. D \textbf{102}, 014005 (2020).

\bibitem{Brodsky:1994eh}
  S.~J.~Brodsky and H.~J.~Lu,
  Commensurate scale relations in quantum chromodynamics,
  Phys.\ Rev.\ D {\bf 51}, 3652 (1995).

\bibitem{Lu:1992nt}
  H.~J.~Lu and S.~J.~Brodsky,
  Relating physical observables in QCD without scale - scheme ambiguity,
  Phys.\ Rev.\ D {\bf 48}, 3310 (1993).

\bibitem{Hoang:1995ex}
  A.~H.~Hoang, J.~H.~Kuhn and T.~Teubner,
  Radiation of light fermions in heavy fermion production,
  Nucl.\ Phys.\ B {\bf 452}, 173 (1995).

\bibitem{Hoang:1997sj}
  A.~H.~Hoang,
  Two loop corrections to the electromagnetic vertex for energies close to threshold,
  Phys.\ Rev.\ D {\bf 56}, 7276 (1997).

\bibitem{Czarnecki:1998qc}
  A.~Czarnecki and K.~Melnikov,
  Two loop QCD corrections to top quark width,
  Nucl. Phys. B \textbf{544}, 520 (1999).

\bibitem{Chetyrkin:1999ju}
  K.~G.~Chetyrkin, R.~Harlander, T.~Seidensticker and M.~Steinhauser,
  Second order QCD corrections to $t \to W^{+} b$,
  Phys. Rev. D \textbf{60}, 114015 (1999).

\bibitem{Blokland:2004ye}
  I.~R.~Blokland, A.~Czarnecki, M.~Slusarczyk and F.~Tkachov,
  Heavy to light decays with a two loop accuracy,
  Phys. Rev. Lett. \textbf{93}, 062001 (2004).

\bibitem{Blokland:2005vq}
  I.~R.~Blokland, A.~Czarnecki, M.~Slusarczyk and F.~Tkachov,
  Next-to-next-to-leading order calculations for heavy-to-light decays,
  Phys. Rev. D \textbf{71}, 054004 (2005).

\bibitem{Gao:2012ja}
  J.~Gao, C.~S.~Li and H.~X.~Zhu,
  Top Quark Decay at Next-to-Next-to Leading Order in QCD,
  Phys. Rev. Lett. \textbf{110}, 042001 (2013).

\bibitem{Brucherseifer:2013iv}
  M.~Brucherseifer, F.~Caola and K.~Melnikov,
  $\mathcal O(\alpha_s^2)$ corrections to fully-differential top quark decays,
  JHEP \textbf{04}, 059 (2013).

\bibitem{CMS:2014mxl}
  V.~Khachatryan \textit{et al.} [CMS],
  Measurement of the ratio $\mathcal B(t \to Wb)/\mathcal B(t \to Wq)$ in pp collisions at $\sqrt{s}$ = 8 TeV,
  Phys. Lett. B \textbf{736}, 33-57 (2014).

\bibitem{Basdevant:1972fe}
  J.~L.~Basdevant,
  The Pade approximation and its physical applications,
  Fortsch. Phys. \textbf{20}, 283 (1972).

\bibitem{Samuel:1992qg}
  M.~A.~Samuel, G.~Li and E.~Steinfelds,
  Estimating perturbative coefficients in quantum field theory using Pade approximants. 2.,
  Phys. Lett. B \textbf{323}, 188 (1994).

\bibitem{Samuel:1995jc}
  M.~A.~Samuel, J.~R.~Ellis and M.~Karliner,
  Comparison of the Pade approximation approach to perturbative QCD calculations,
  Phys. Rev. Lett. \textbf{74}, 4380 (1995).

\bibitem{Cacciari:2011ze}
  M.~Cacciari and N.~Houdeau,
  Meaningful characterisation of perturbative theoretical uncertainties,
  JHEP \textbf{09}, 039 (2011).

\bibitem{Bagnaschi:2014wea}
  E.~Bagnaschi, M.~Cacciari, A.~Guffanti and L.~Jenniches,
  An extensive survey of the estimation of uncertainties from missing higher orders in perturbative calculations,
  JHEP \textbf{02}, 133 (2015).

\bibitem{Bonvini:2020xeo}
  M.~Bonvini,
  Probabilistic definition of the perturbative theoretical uncertainty from missing higher orders,
  Eur. Phys. J. C \textbf{80}, 989 (2020).

\bibitem{Duhr:2021mfd}
  C.~Duhr, A.~Huss, A.~Mazeliauskas and R.~Szafron,
  An analysis of Bayesian estimates for missing higher orders in perturbative calculations,
  JHEP \textbf{09}, 122 (2021).

\bibitem{Gardi:1996iq}
  E.~Gardi,
  Why Pade approximants reduce the renormalization scale dependence in QFT?,
  Phys. Rev. D \textbf{56}, 68 (1997).

\bibitem{Cvetic:1997qm}
  G.~Cvetic,
  Improvement of the approach of diagonal Pade approximants for perturbative series in gauge theories,
  Phys. Rev. D \textbf{57}, R3209 (1998).

\bibitem{Du:2018dma}
  B.~L.~Du, X.~G.~Wu, J.~M.~Shen and S.~J.~Brodsky,
  Extending the Predictive Power of Perturbative QCD,
  Eur. Phys. J. C \textbf{79}, 182 (2019).

\bibitem{Shen:2022nyr}
  J.~M.~Shen, Z.~J.~Zhou, S.~Q.~Wang, J.~Yan, Z.~F.~Wu, X.~G.~Wu and S.~J.~Brodsky,
  Extending the Predictive Power of Perturbative QCD Using the Principle of Maximum Conformality and Bayesian Analysis,
  [arXiv:2209.03546 [hep-ph]].

\end{thebibliography}
\end{document}